\documentclass[aps,10pt,pof,notitlepage,onecolumn,groupedaddress,longbibliography]{revtex4-1}
\usepackage{dcolumn,bm,hyperref,lmodern,units,mdwlist,graphicx}
\usepackage{multirow,url,setspace,natbib,graphics,amssymb,amsmath,float,upgreek}
\usepackage{subfig,textcomp,listings,psfrag,placeins,epsfig,caption,colortbl}
\usepackage{caption,subfig}
\usepackage[toc,page]{appendix}
\newcommand{\R}{\mathcal{R}}
\newcommand{\Pt}{\mathcal{P}_{sp,t}}
\newcommand{\Pl}{\mathcal{P}_{sp,l}}
\newcommand{\Px}{\mathcal{P}_{x}}
\newcommand{\C}{\mathcal{C}}
\newcommand{\Sx}{\mathcal{S}_x}
\newcommand{\Sz}{\mathcal{S}_z}
\newcommand{\Rt}{R_\tau}
\newcommand{\Rp}{R_p}
\newcommand{\p}{\partial}
\newcommand{\E}{\mathcal{E}}
\newcommand\numberthis{\addtocounter{equation}{1}\tag{\theequation}}

\begin{document}

\title{Spinning out of control: wall turbulence over rotating discs}
\author{Daniel J. Wise}
\email[Email:]{d.wise@sheffield.ac.uk}
\author{Claudia Alvarenga}
\author{Pierre Ricco}
\affiliation{Department of Mechanical Engineering, The University of Sheffield,\\
Mappin Street, S1 3JD Sheffield, United Kingdom}

\begin{abstract}
The friction drag reduction in a turbulent channel flow generated by surface-mounted rotating disc actuators is investigated numerically.
The wall arrangement of the discs has a complex and unexpected effect on the flow.  For low disc-tip velocities, the drag reduction scales linearly with the percentage of the actuated area, whereas for higher disc-tip velocity the drag reduction can be larger than the prediction found through the linear scaling with actuated area. For medium disc-tip velocities, all the cases which display this additional drag reduction exhibit stationary-wall regions between discs along the streamwise direction. This effect is caused by the viscous boundary layer which develops over the portions of stationary wall due to the radial flow produced by the discs. For the highest disc-tip velocity, the drag reduction even increases by halving the number of discs.  The power spent to activate the discs is instead independent of the disc arrangement and scales linearly with the actuated area for all disc-tip velocities. The Fukagata-Iwamoto-Kasagi identity and flow visualizations are employed to provide further insight into the dynamics of the streamwise-elongated structures appearing between discs. Sufficient interaction between adjacent discs along the spanwise direction must occur for the structures to be created at the disc side where the wall velocity is directed in the opposite direction to the streamwise mean flow.  Novel half-disc and annular actuators are investigated to improve the disc-flow performance, resulting in a maximum of 26\% drag reduction.
\end{abstract}

\maketitle

\section{\label{sec:introduction}Introduction}
Turbulent skin-friction drag reduction has been the subject of growing interest in the fluid mechanics research community in recent decades.  A breakthrough in this context would lead to lower fuel consumption and improved ecosustainability in many industrial scenarios, and it is for this reason that great efforts are directed towards improving the understanding of the underlying physical mechanisms and to the development of novel drag reduction techniques.

Flow control techniques can be classified as active or passive.  Active methods are those which require an external energy input, while passive methods manipulate the flow field without a supply of energy.  Amongst active methods there exists a further division between techniques which operate under closed- or open-loop control \citep{gadelhak-2000}.  Closed-loop control requires sensors to measure the flow properties, thus allowing the control input to be adjusted according to a prescribed algorithm. Open-loop control is instead predetermined and does not respond to changes in the flow.  As such it does not require sensors.  Although numerical investigations of closed-loop flow control utilizing linear control theory have promised high drag reduction and significant net power savings (computed by taking into account the energetic cost of control), the experimental verification of these computational efforts poses enormous challenges.  These relate to the very small spatial and temporal scales typically required to achieve such energetic performances.  Progress is nonetheless being made with the fabrication of novel MEMS-based flow sensors and actuators \citep{kasagi-suzuki-fukagata-2009}.  According to the estimates of \citet{wilkinson-1990} the current production cost of such systems for use on a commercial aircraft would however render their application prohibitively expensive.

Promisingly, active open-loop control reaches a compromise between complexity and performance.  Since the pioneering direct numerical simulations (DNS) of \citet{jung-mangiavacchi-akhavan-1992} and the experiments of \citet{laadhari-skandaji-morel-1994}, the response of wall-bounded turbulent flows to spanwise, spatially uniform sinusoidal oscillations of the wall has become one of the most studied active open-loop techniques. The temporal forcing has been converted to spatial forcing in the form of standing waves and has been confirmed to produce wall-friction reductions of up to 40\% \citep{viotti-quadrio-luchini-2009}.  Drag reduction is thought to occur because the intensity of the Reynolds stresses decreases as a result of the weakening of the turbulence structures \citep{choi-xu-sung-2002}.  \citet{skote-2011} employed the steady waves to alter a streamwise-developing boundary layer and observed strong suppression of low-speed streaks above those parts of the wall for which the velocity was maximum.  Furthermore, \citet{skote-2013} showed that the improved drag reduction for spatial oscillations over temporal oscillations may be explained by an additional negative turbulence production term involving the streamwise gradient of the spanwise velocity.  A generalization of the oscillating-wall and standing-wave forcing was proposed by \citet{quadrio-ricco-viotti-2009}, who studied the response of a turbulent channel flow to streamwise travelling waves of spanwise wall velocity.  They showed a maximum drag reduction of 47\% and a maximum net energy saving of 26\%.  However, it remains to be shown whether techniques such as these, which involve large scale motions of the entire wall and short time scales, will become attractive for industrial applications.  The experimental works of \citet{gouder-potter-morrison-2013} on electroactive polymers and of \citet{choi-jukes-whalley-2011} on dielectric-barrier discharge plasma actuators are certainly advances in this respect.

Another example which represents a further step towards application is the actuation strategy first proposed by \citet{keefe-1998}, based on arrays of flush-mounted discs rotating in response to the detection of the turbulent bursting process.  Despite the promising outlook on the applicability of this technique and the prediction of the optimal disc diameter and rotation frequency ($80-90\upmu\mbox{m}$ and $72\mbox{kHz}$ respectively), Keefe did not further investigate his idea, and in the following 15 years neither experimental nor numerical studies on this flow appeared.  \citet{ricco-hahn-2013} (denoted by RH13 hereafter) were the first to follow up with a numerical investigation of the disc actuators, whereby the discs rotated with a constant angular velocity.  A parametric investigation on $D$, the disc diameter, and on $W$, the disc-tip velocity, yielded maximum drag reduction and net power savings of 23\% and 10\%, respectively.  Flow visualizations unexpectedly revealed the existence of streamwise-elongated tubular structures between the discs.  Through the use of the Fukagata-Iwamoto-Kasagi identity\citep{fukagata-iwamoto-kasagi-2002} (FIK), RH13 showed that the Reynolds stresses associated with these structures contribute favourably to the overall drag reduction effect.  It was further shown that the power spent is satisfactorily predicted by the laminar solution for the flow over an infinite rotating disc and that drag reduction occurs only when the boundary layer engendered by the disc rotation is thicker than a threshold.  Furthermore, drag increase was computed in a range of small $D$ and high $W$.

Flows over rotating discs have been studied extensively, beginning with the exact similarity solution to the flow over an infinite spinning disc given by \citet{karman-1921}.  The first numerical results on this flow were obtained by \citet{cochran-1934}.  These works were extended by \citet{rogers-lance-1960} to include solutions to the flow induced by a disc rotating beneath a swirling fluid. The similarity solution to the case of a rotating plate beneath a streamwise laminar shear flow was first determined by \citet{wang-1989}.  He showed that the presence of an external flow caused a streamwise shift in the stagnation point on the disc. \citet{klewicki-hill-2003} first experimentally investigated the response of a laminar boundary layer to the rotation of a surface patch, observing results consistent with the Wang solution.  Other prominent studies on rotating disc flows include the theoretical and experimental stability analyses by \citeauthor{lingwood-1995} \citep{lingwood-1995,lingwood-1996,lingwood-1997b}.  The results presented in this paper complement this list and extend the line of research on wall turbulence modified by flush-mounted discs, first explored by RH13 and \citet{wise-ricco-2014} (denoted by WR14 hereafter).

The aim of the current work is to provide further insight into the rotating disc technique of RH13.  Direct numerical simulations of a turbulent channel flow are employed to investigate the effects of different disc layouts on the drag reduction, the power spent, and the interdisc structures.  The influence on wall turbulence of rotating annular discs and of the configuration of RH13 with the downstream half of the discs covered by a solid wall is investigated. The flow response to disc actuation for which only part of the spectral distribution of the wall velocity is actuated is also studied.  We close this paper with an appendix discussing the prediction of the power spent via the laminar flow induced by the disc motion below a quiescent fluid.  The focus in this appendix is on the steady rotation case and on the oscillating case, studied by WR14.  It has recently occurred to us that the mathematical expressions derived in those publications pertain to the power spent per unit of activated area, i.e. where the wall velocity is non-zero.  In order to have a meaningful comparison with the power spent computed via DNS, the laminar power spent is derived by averaging over the whole wetted wall area.  The prediction is further improved by modelling the effect of the clearance around the discs on the power spent.

The numerical procedures, disc arrangements, averaging procedures, flow decompositions, and definitions of performance quantities are found in Sec.~\ref{sec:methods}.  The effect of layout and coverage on the performance quantities is outlined in Sec.~\ref{sec:layout}.  The FIK identity is employed to investigate the flow in Sec.~\ref{sec:fik} and flow visualizations are studied in Sec.~\ref{sec:flowvis}.  A discussion of the radial flow induced by the discs is contained in Sec.~\ref{sec:radial}.  Modifications to the actuators to improve the drag reduction effect are presented in Sec.~\ref{sec:hd} and Sec.~\ref{sec:ann}. The influence of large and small scale forcing on the performance quantities is investigated in Sec.~\ref{sec:spect}. Sec.~\ref{sec:summary} presents a summary of the results.  Appendix~\ref{app:drresults} contains a table of the drag reduction and power spent data. Appendix~\ref{app:psp} outlines the power spent predictions via the laminar flow solution and includes corrections to the formulae given in RH13 and WR14.

\section{\label{sec:methods}Numerical procedures}
\subsection{\label{sec:numerics}Numerical solver, geometry and scaling}
\begin{figure}
\centering
\includegraphics[width=0.75\textwidth]{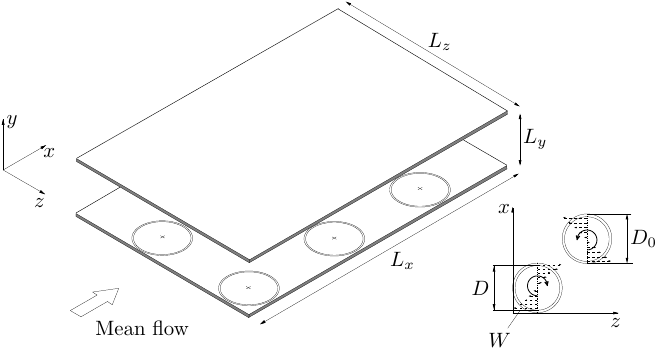}
\caption{Schematic of simulated channel geometry. The disc layout shown is case 4 (refer to Fig.~\ref{fig:arrangements} for other layouts).}
\label{fig:rh13discs}
\end{figure}

A pressure-driven turbulent channel flow at constant mass flow rate is investigated by DNS.  The infinite, parallel flat walls of the channel are separated by $L^*_y$=$2h^*$.  The symbol $^*$ denotes a dimensional quantity.  A schematic of the flow domain is shown in Fig.~\ref{fig:rh13discs}.  $L_x^*$ and $L_z^*$ are the dimensions of the computational domain in the streamwise ($x^*$) and spanwise ($z^*$) directions.  Simulations are performed at $\Rp$=$U_p^*h^*/\nu^*$=4200, where $\nu^*$ is the kinematic viscosity of the fluid and $U_p^*$ is the centreline velocity of the laminar Poiseuille flow at the same mass flow rate. The equivalent friction Reynolds number in the fixed-wall configuration is $\Rt$=$u_\tau^*h^*/\nu^*$=180, where $u_\tau^*$=$\sqrt{\tau^*/\rho^*}$ is the friction velocity, $\tau^*$ is the space- and time-averaged wall-shear stress, and $\rho^*$ is the density. An open-source code, available on the Internet \cite{channelflow-2006}, is utilized to solve the incompressible Navier-Stokes equations using Fourier series expansions along the statistically homogeneous $x^*$ and $z^*$ directions, and Chebyshev polynomials along the wall-normal direction $y^*$.  A third-order semi-implicit backward differentiation scheme is used to advance the equations in time. The discretized equations are solved using the Kleiser-Schumann algorithm \citep{kleiser-schumann-1980}, described in \citet{canuto-etal-1988}. The nonlinear terms are treated explicitly and the linear terms implicitly.  Dealiasing is carried out by setting the upper third of the modes in the $x$ and $z$ directions to zero.  The wall boundary conditions were modified by RH13 to implement the disc motion.  The code is parallelized using OpenMP and simulations have been carried out on the N8 HPC Polaris cluster.  Post-processing has been performed on the Iceberg cluster at the University of Sheffield.

Lengths are scaled by $h^*$, velocities by $U_p^*$, and time by $h^*/U_p^*$. Scaling using these outer units is not marked by any symbol.  Quantities denoted by the $+$ superscript are scaled in viscous units, i.e. with $\nu^*$ and $u_\tau^*$, where $u_\tau^*$ pertains to the uncontrolled reference case. For $D$=3.38, the size of the computational domain is ($L_x$,$L_y$,$L_z$)=(4.52$\pi$,2,2.26$\pi$) and, for $D$=5.02, ($L_x$,$L_y$,$L_z$)=(6.79$\pi$,2,3.39$\pi$). The resolution along $x$ and $z$ is constant in all cases, $\Delta x^+$=10 and $\Delta z^+$=5, corresponding to a number of Fourier modes equal to $N_x$=$N_z$=256 for $D$=3.34 and $N_x$=$N_z$=384 for $D$=5.02. The number of grid points in the wall-normal direction is kept constant at $N_y$=129. Nodes along $y$ are clustered according to $y(i)$=$\cos\left[i\pi/(N_y-1)\right]$, where 0$\leq\hspace{-0.15cm}i$\textless$N_y$, $\Delta y_{min}$=0.054, and $\Delta y_{max}$=4.42.  This allows greater resolution near the walls. The time step is changed adaptively between $\Delta t^+_{min}$=0.008 and $\Delta t^+_{max}$=0.08. This reduces the computational cost by maximizing the CFL number within the range 0.2$<$CFL$<$0.4.

\subsection{Arrangement of discs}
\begin{figure}
\centering
\includegraphics[width=0.65\textwidth]{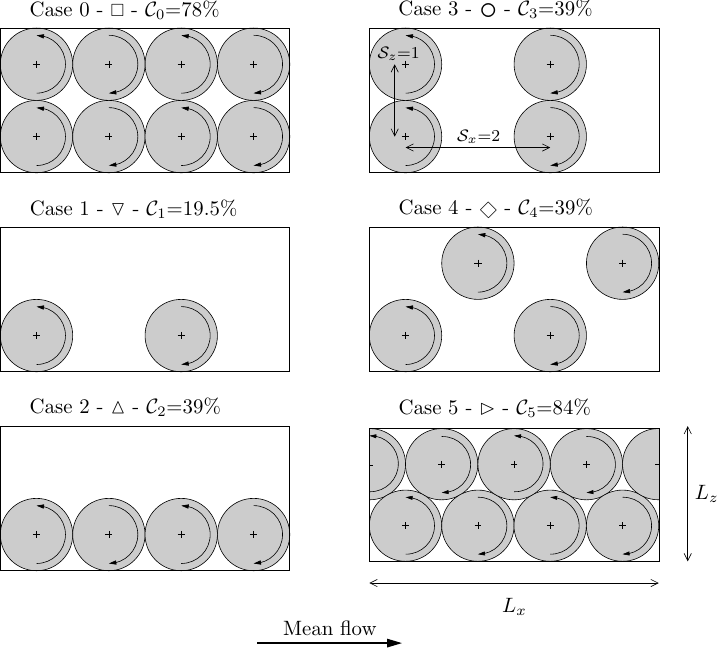}
\caption{Disc layouts in the wall $x$$-$$z$ plane.}
\label{fig:arrangements}
\end{figure}

The discs are located on both walls, have diameter $D$ and rotate steadily with an angular velocity $\Omega$. The disc-tip velocity is $W$=$\Omega D/2$. In RH13 the discs are arranged in a square packing scheme, with discs which are adjacent in the streamwise direction spinning in opposite directions and discs along the spanwise direction rotating in the same direction.  This configuration was chosen to resemble the standing wave studied by \citet{viotti-quadrio-luchini-2009}, and will henceforth be referred to as case 0. The layout for case 0 and the modified disc arrangements investigated herein are presented in Fig.~\ref{fig:arrangements}. The coverage $\C$ is defined as the percentage of the wall surface which is in motion. For each arrangement, a coverage $\C_n$ is defined, with the subscript $n$ referring to the layouts as numbered in Fig.~\ref{fig:arrangements}. For the reference case studied by RH13 (case 0), $\C_0$=78\%. For case 5, the arrangement is not the hexagonal lattice that gives maximum coverage for packing of equal circles (i.e. $\C$=91\%).  As the channel domain must be rectangular, it is not possible to configure the discs in this manner whilst maintaining an integer number of discs. The layout shown at the bottom right of Fig.~\ref{fig:arrangements} is instead simulated.  The coverage for this arrangement is $\C_5$=84\% and an integer number of discs is enforced. The spanwise length of the domain for case 5 is $L_z$=2.11$\pi$ for $D$=3.38 and $L_z$=3.17$\pi$ for $D$=5.02, due to the hexagonal disc arrangement.

The disc diameters and velocities studied are $D$=3.38 and 5.02, and $W$=0.13,0.26,0.39, and 0.52.  These forcing parameters are the ones that guarantee a high drag reduction of about 20\% in the configuration studied by RH13.  The term column is used to indicate disc alignment along the streamwise direction and the term row is used to denote disc alignment along the spanwise direction.

\subsection{Averaging procedures and flow decomposition}

The time average is defined as
\begin{equation*}
\overline{{\bf f}}(x,y,z) = \frac{1}{t_f-t_i}\int_{t_i}^{t_f} {\bf f}(x,y,z,t) \textup{d}t,
\label{average-t}
\end{equation*}
where $t_i$ and $t_f$ denote the start and finish of the averaging time.  The spatial average along the homogeneous directions is defined as
\begin{equation*}
\langle{\bf f}(y)\rangle = \frac{1}{L_xL_z}\int_0^{L_x}\int_0^{L_z}\overline{{\bf f}}(x,y,z)\textup{d}z\textup{d}x.
\end{equation*}
The flow field within the channel is expressed as the sum of three components,
\begin{equation}
{\bf u}={\bf u_m}+{\bf u_d}+{\bf u_t}\text{,}
\label{eq:decomp}
\end{equation}
where ${\bf u_m}(y)$=$\left\{u_m(y),0,0\right\}$=$\langle{\bf u}\rangle$ is the mean flow, ${\bf u_d}(x,y,z)$=$\left\{u_d,v_d,w_d\right\}$=$\overline{{\bf u}}$$-$${\bf u_m}$ is the disc flow, and ${\bf u_t}$ represents the turbulent fluctuations.  Flow fields have been computed over a minimum integration time of $1400h^*/U_p^*$.  This time window does not include the initial transient from the start of the disc motion, during which the flow adjusts to the new forcing conditions.  All statistical samples are doubled by averaging over both halves of the channel, by accounting for the existing symmetries with respect to the centreline of the channel.
\subsection{\label{sec:pq}Performance quantities}
The turbulent drag reduction is defined as
\begin{equation}
\R(\%)=100\frac{C_{f,s}-C_f}{C_{f,s}}\text{,}
\label{eq:rcf}
\end{equation}
where $C_f$=2$\tau^*/(\rho^*U_b^{*2})$ is the skin-friction coefficient, $U_b$=$\int_0^1u_m(y)\textup{d}y$ is the bulk velocity, and the subscript $s$ denotes the stationary-wall case.  Since simulations are carried out under constant mass flow rate conditions, $U_b$=2/3 throughout.  As shown by RH13, the power supplied to the discs to rotate them against the viscous resistance of the fluid and expressed as a percentage of the power needed to pump the fluid in the streamwise direction, is
\begin{equation*}
\Pt(\%)=\frac{100\Rp}{2\Rt^2U_b}
\left.
\frac{\textup{d}\hspace{-1mm}\left(u^2_d+w_d^2\right)}{\textup{d}y}
\right|_{y=0}\text{.}
\end{equation*}
\subsection{\label{sec:ann-gap}Annular gap}
\begin{figure}%
\centering
  \includegraphics[width=0.45\textwidth]{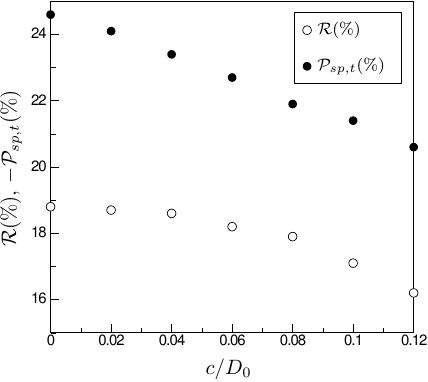}
  \caption{Drag reduction $\R$ and power spent $\Pt$ vs. $c/D_0$ for $D_0$=3.56 and $W$=0.39.}
  \label{fig:gaptable}%
\end{figure}

As in RH13, a small annular region of thickness $c$ is simulated around each disc.  The wall velocity in this region decays linearly from the maximum at the disc tip to zero at the stationary wall and is independent from the azimuthal direction. The azimuthal velocity $u_\theta$ varies with the radial coordinate $r$ as follows:
\begin{equation*}
  u_\theta(r) = \left\{
  \begin{array}{l l}
    2Wr/D,
    &
r \leq D/2\text{,} \\
    W(c-r+D/2)/c,
    &
D/2 \leq r \leq D/2+c\text{.}\\
  \end{array} \right.
\end{equation*}
This serves to mimic an experimental scenario where a gap would inevitably be present. As shown by RH13, the Gibbs phenomenon at the disc edges is also almost entirely suppressed. It would be significant if the gap were not simulated because of the velocity discontinuity at the boundary between the disc tip and stationary wall. The effect of gap size on the performance quantities for $D_0$=3.56 and $W$=0.39 is shown in Fig.~\ref{fig:gaptable}, where $D_0$=$D$$+$$2c$ is the outer diameter of the circle occupied by the disc and the annular gap, as shown in Fig.~\ref{fig:rh13discs}. Although the Gibbs phenomenon does occur for $c$=0, it does not influence the computation of drag reduction as the effect is limited to the disc edge. The drag reduction decreases by about 1\% as $c$ increases from 0 to 0.08$D_0$.  It then decreases more rapidly and, by $c$=0.12$D_0$, $\R$ is 70\% of the value obtained without the annular gap.  The power spent decreases almost linearly and more rapidly than $\R$ as the gap size increases.  The averaged wall-shear stress therefore responds primarily to the large scales of the disc forcing, while the power spent shows a more marked dependence on the precise distribution of wall actuation.  More evidence of this emerges in Sec.~\ref{sec:spect} where the dependence of these quantities on the spectral representation of wall forcing is investigated.  The gap size in the following cases is $c$=0.06$D_0$, which would most closely resemble the clearance in a water channel or in a wind tunnel set up.

The drag reduction computed in RH13 for $D$=3.38, $W$=0.39, and $c/D_0$=0.05 is $\R$=19.5\%, which is larger than the corresponding value estimated from the data in Fig.~\ref{fig:gaptable}, $\R$=18.5\%.  This discrepancy is larger than the uncertainty range of the numerical calculations.  The difference between the $C_f$ in the actuated-wall case in RH13 ($C_f$=6.64$\cdot10^{-3}$) and the $C_f$ computed here for $c/D_0$=0.06 ($C_f$=6.68$\cdot10^{-3}$) leads to only a 0.4\% difference in $\R$ if the stationary-wall $C_f$ computed by RH13 is used as reference case ($C_f$=8.25$\cdot10^{-3}$).  More accurate resolution checks on the stationary-wall $C_f$ lead to $C_f$=8.19$\cdot10^{-3}$, which explains the 1\% difference in $\R$.

\section{\label{sec:results}Results and discussion}
\subsection{\label{sec:layout}Influence of layout and coverage}
\subsubsection{Drag reduction}
\begin{figure}
\centering
\includegraphics[width=0.85\textwidth]{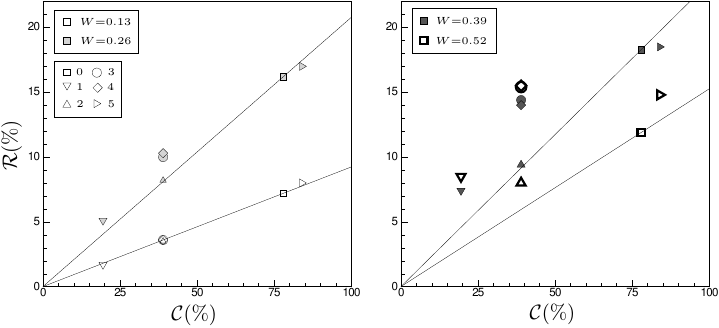}
\caption{Drag reduction vs. coverage for $D$=3.38. In the legend, the symbols are numbered according to the layouts in Fig.~\ref{fig:arrangements} and are coloured according to $W$.}
\label{fig:r-cov}
\end{figure}

\begin{figure}
\centering
\includegraphics[width=0.425\textwidth]{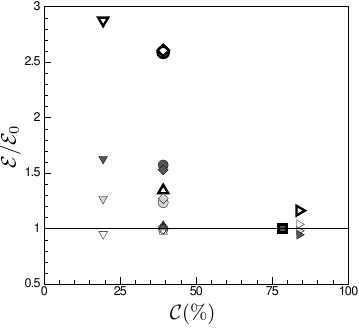}
\caption{Drag reduction gain $\E/\E_0$ vs. coverage for $D$=3.38.  Symbols are as in Fig.~\ref{fig:arrangements} and coloured according to the legends shown in Fig.~\ref{fig:r-cov}.}
\label{fig:norm-r-over-c}
\end{figure}

The drag reduction $\R$ is shown in Figs.~\ref{fig:r-cov} and \ref{fig:norm-r-over-c} as a function of the coverage $\C$ for $D$=3.38 and different $W$. The numerical values are found in Appendix \ref{app:drresults}.  The different symbols denote the different arrangements and the different colours indicate different $W$. The solid lines in Fig.~\ref{fig:r-cov} represent the drag reduction predicted through $\R$=$(\C/\C_0)\R_0$, i.e. via straight lines passing through the origin and the $\R_0$ values by RH13. These are not interpolating lines of the drag reduction data. $\R$ values falling on these lines obey linear scaling with coverage.  For cases with $W$=0.13, shown by the white symbols, $\R$ scales linearly with $\C$.  This implies that the drag reduction is only produced by the shearing effect of the flow over the disc surface.  The hexagonal arrangement (case 5), which gives the maximum wall coverage $\C_5$=84\%, also follows the linear scaling with $\C$.  The scaling starts to deteriorate for some of the cases with $W$=0.26 and 0.39 (light and dark grey symbols), and is completely lost for $W$=0.52 (bold white symbols).  A different physical mechanism must be responsible for drag reduction for the cases which do not follow the linear scaling with coverage.  Except for case 5 and $W$=0.39, in all the cases that do not fall on the straight lines, $\R$ is larger than the corresponding value predicted by the coverage scaling. The drag reduction for case 0 and $W$=0.52 ($\R$=11.9\%) is lower than the one given by cases 3 and 4 for the same $W$ and $D$ ($\R$=15.5\%) despite the removal of half of the discs.

For cases with $\C_1$=19.5\%, in which the surface is covered by a fourth of the number of discs used by RH13, the additional drag reduction with respect to coverage increases monotonically with $W$. Although cases 2, 3, and 4 all have the same coverage, $\C$=39\%, the drag reduction values differ for the same $W$ and $D$ because they have different disc arrangements.  Case 2, for which discs are aligned in one column (upward facing triangles), obeys coverage scaling up to $W$=0.39.  Case 3, for which discs aligned along every other row (circles), and case 4, which has a checkerboard disc arrangement (diamonds), instead lose this scaling for $W\hspace{-0.1cm}\geq$0.26. At the same $W$, the $\R$ values of cases 2 and 3 only differ by small amounts, which are within the uncertainty range for all the $W$ tested.  For 0.26$\leq\hspace{-0.1cm}W\hspace{-0.1cm}\leq$0.39, it follows that the additional drag reduction with respect to the value predicted by the linear scaling with coverage occurs when a portion of stationary wall of the streamwise extent of one diameter is present between discs. The spanwise space between discs does not have an effect because case 3 (discs next to each other along $z$) and case 4 (spanwise space at either side of discs) lead to the same drag reduction.

The case of hexagonal arrangement, $\C_5$=84\%, presents drag reduction values which are shifted below the coverage line for $W$=0.39.  This is consistent with the upward shift of cases which present a streamwise region of stationary wall. In the hexagonal arrangement the streamwise spacing between discs is instead reduced and therefore drag reduction deteriorates with respect to the coverage line.

The drag reduction given by case 2 (discs aligned in one column) loses the linear scaling only at $W$=0.52, even though no streamwise spacing is present.  An upward shift with respect to the coverage line also occurs for case 5 at $W$=0.52.  Similarly to the upward shift of case 2 at the same $W$, this is not due to the streamwise fixed-wall space as in cases 1, 3, and 4 because discs are closely packed along the streamwise direction.  It is neither due to the spanwise space of fixed wall at the side of each disc because the additional drag reduction is the same in cases 2 and 5, although case 2 displays more spanwise space than case 5.  The drag reduction at $W$=0.52 being higher than the value predicted by the linear scaling with coverage remains unexplained at this point.

By defining a new quantity, $\E$=$\R/\C$, the coverage gain of the disc actuators is given as the drag reduction induced per actuated area.  For cases in which $\E$\textgreater$\E_0$, where $\E_0$=$\R_0/\C_0$ is the coverage gain for case 0, larger drag reduction occurs compared to case 0 for the same number of discs.  Fig.~\ref{fig:norm-r-over-c} presents $\E/\E_0$ as a function of $\C$. In this scaling, it emerges that the gain is null at $W$=0.13, independent of $\C$ when $W$=0.26 for cases that do not follow coverage, and at its maximum at low coverage and high $W$.

\begin{figure}
\centering
\subfloat{\includegraphics[width=0.425\textwidth]{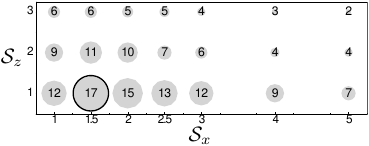}}\quad
\subfloat{\raisebox{1.8cm}{
  \begin{tabular}[c]{c|cc|c|c|c}
  Case	&	$D$	&	$W$	&	$\R_{sw}(\%)$	&	$\R_{pred}(\%)$	&	$\R(\%)$	\\ \hline
    0	& 	3.38   	& 	0.26	&	33		&	16.4		&   	16.2		\\
    2	& 	3.38   	& 	0.26	&	33		&	8.2		&   	8.2		\\
    5	& 	3.38   	& 	0.13	&   	16		&	8.6		&	8.7		\\
    \end{tabular}
    }
}
\caption{Left: Map of $\mathcal{R}(\Sx,\Sz)(\%)$ for $D$=3.38, $W$=0.52.  Right: Comparison of drag reduction data from the DNS with those given from rescaling of \citet{viotti-quadrio-luchini-2009}.}
\label{fig:spacing}
\end{figure}

For the cases examined heretofore, the displacement between adjacent streamwise and spanwise disc centres has been either $D_0$ or 2$D_0$.  More arrangements of discs can be studied by defining the spacings $\Sx$=$x_d/D_0$ and $\Sz$=$z_d/D_0$, where $x_d$ and $z_d$ are the distances between neighbouring disc centres in the $x$ and $z$ directions, respectively.  $\Sx$ and $\Sz$ are shown graphically in case 3 in Fig.~\ref{fig:arrangements}.  Fig.~\ref{fig:spacing} (left) shows $\R$ for different $\Sx$ and $\Sz$ with disc parameters $D$=3.38, $W$=0.52.  An optimum spacing is found for $(\Sx,\Sz)$=(1.5,1) resulting in $\R$=17\%. For comparison the RH13 value (case 0) is $\R$=12\% for the same disc parameters.

As $\R$ scales with coverage at low $W$, a prediction of the drag reduction engendered by the discs is attempted, starting from the data computed in \citet{viotti-quadrio-luchini-2009} (page 10) for the standing-wave case.  As noted by RH13, the wall forcing created along the disc centres is similar to a triangular wave of wavelength $\lambda_x$=2$D_0$ and amplitude $W$.  The drag reduction given by the discs can be predicted as $\R_{pred}$=$C_w\cdot C_\theta\cdot\C\cdot\R_{sw}$, where $C_w$ is the scaling factor due to waveform, $C_\theta$ models the effect of the orientation of wall forcing, $\C$ accounts for the wall coverage, and $\R_{sw}$ is the drag reduction in the standing-wave case by \citet{viotti-quadrio-luchini-2009} for $\lambda_x$=2$D_0$. The factors are approximated as follows.

\noindent
\begin{description*}
\item[\it Waveform] It is known that temporal and spatial forcing can be largely treated as analogous to one another \citep{quadrio-ricco-viotti-2009}. The temporal non-sinusoidal spanwise wall-forcing investigated by \citet{cimarelli-etal-2013} can thus be used to gauge the influence of the spatially non-sinusoidal spanwise wall-forcing of the discs.  Waveform {\em j} on page 4 of \citet{cimarelli-etal-2013} closely resembles the triangular wave spanwise forcing of the discs, which results in $C_w$=85\%.
\item[\it Streamwise forcing] The streamwise forcing which is present in the disc technique does not occur in the standing-wave case studied by \citet{viotti-quadrio-luchini-2009}. The effect of wall oscillations at an angle $\theta$ with respect to the mean flow has been studied by \citet{zhou-ball-2006}. While pure spanwise oscillations produce the maximum drag reduction, the response to streamwise oscillations reduces to a third.  The influence of wall-forcing orientation is accounted for by $C_\theta$=75\%, estimated by averaging \citeauthor{zhou-ball-2006}'s data over the angle of wall forcing.
\item[\it Coverage] This is quantified by the coverage value $\C_n$ for each case, given in Fig.~\ref{fig:arrangements}.
\end{description*}

\noindent
The table in Fig.~\ref{fig:spacing} (right) shows the $\R$ values for three sample layouts and disc parameter combinations.  The prediction $\R_{pred}$ of the numerically computed $\R$ is excellent for the cases tested.

\subsubsection{Power spent}
The effect of coverage is now studied on the power spent, shown as a function of $\C$ in Fig.~\ref{fig:p-cov}.  The numerical values are found in Appendix \ref{app:drresults}.  For all $W$ the linear scaling of power spent with coverage is excellent and much more robust than for drag reduction, shown in Fig.~\ref{fig:r-cov}. The power spent therefore does not depend on the disc arrangements for fixed $\C$.  This follows from the power spent being solely related to the wall motion and largely independent of the dynamics of turbulence within the channel.  The solid lines represent the laminar prediction to the power spent $\Pl$, calculated from the solution to the flow induced by an infinite disc rotating beneath a quiescent fluid \citep{batchelor-1967}. An amended and improved version of the formula in RH13, which now takes into account the effect of the gap flow, is derived in Appendix \ref{app:psp}. It reads
\begin{equation*}
\Pl(\%)=100\frac{\C}{\C_0}
\frac{\pi G_k \Rp^{3/2} W^{5/2}}{U_b\Rt^2 D_0^2}
\sqrt{\frac{2}{D}}
\left(
\frac{D^2}{8}+\frac{cD}{3}+\frac{c^2}{6}
\right)\text{,}
\numberthis
\label{eq:plam}
\end{equation*}
where $G_k$=$-$0.61592 is given in \citet{schlichting-1979}, and $\Rp$ and $\Rt$ are the Poiseuille and friction Reynolds numbers respectively, defined in Sec.~\ref{sec:numerics}.  Equation \eqref{eq:plam} predicts $\Pt$ well, with the turbulent $\Pt$ being always slightly larger than the laminar $\Pl$.
\begin{figure}
\centering
\includegraphics[width=0.425\textwidth]{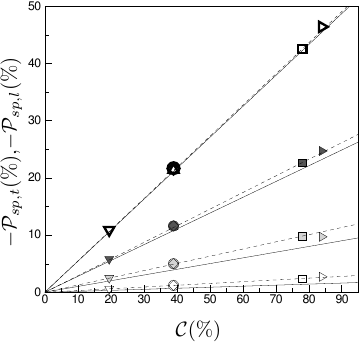}
\caption{Power spent vs. coverage for $D$=3.38.  The different symbols correspond to the different layouts as indicated in Fig.~\ref{fig:arrangements} and are coloured according $W$, as shown in the legend of Fig.~\ref{fig:r-cov}.  The solid lines represent the prediction of power spent by the laminar solution given by \eqref{eq:plam}.  The dashed lines are found by rescaling the RH13 $\Pt$ values with respect to coverage, i.e. they connect the origin and the RH13 values (square symbols).}
\label{fig:p-cov}
\end{figure}

\subsection{\label{sec:fik}The Fukagata-Iwamoto-Kasagi identity}
In this section, the FIK identity \citep{fukagata-iwamoto-kasagi-2002} is used to further understand the mechanism of drag reduction for the disc arrangements studied in Sec.~\ref{sec:layout}. This identity quantifies the effect of the laminar flow and of the Reynolds stresses to the skin-friction coefficient. RH13 and WR14 showed that through this identity it is possible to distinguish two separate contributions to drag reduction, which arise from (a) the modification in the turbulent Reynolds stresses relative to the uncontrolled case, and from (b) the Reynolds stresses $\langle u_dv_d\rangle$, related to the structures appearing between discs and described in RH13 on page 13 and in WR14 on pages 557-558. The drag reduction is written as $\R$=$\R_t$+$\R_d$, where $\R_t$ synthesizes effect (a) and $\R_d$ is related to (b). Their expressions are:
\begin{align*}
\R_t(\%)&=100\frac{\Rp\int_0^1\left(1-y\right)\left(\langle\overline{u_t v_t}\rangle-\langle \overline{u_{t,s}v_{t,s}}\rangle\right)\textup{d}y}{U_b-\Rp\int_0^1\left( 1-y\right)\langle  \overline{u_{t,s}v_{t,s}}\rangle\textup{d}y}\text{,} \\
\R_d(\%)&=100\frac{\Rp\int_0^1\left(1-y\right)\langle u_dv_d\rangle\textup{d}y}{U_b-\Rp\int_0^1\left(1-y\right)\langle\overline{u_{t,s}v_{t,s}}\rangle\textup{d}y}.
\end{align*}

\begin{figure}
\centering
\includegraphics[width=0.425\textwidth]{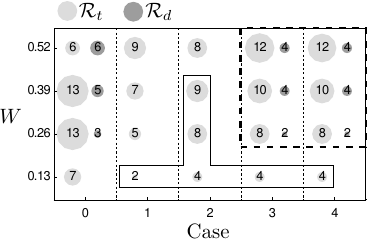}
\caption{Percentage contributions of $\R_t$ and $\R_d$ to $\R$ for $D$=3.38. $\R_t$ represents the contribution from the modification of the turbulent Reynolds stresses (light grey), and $\R_d$ is the contribution from the interdisc structures (dark grey).  Cases boxed by the solid line obey linear scaling with $\C$, while cases boxed by the dashed line are influenced by the interdisc structures.  Case 0 is not boxed because it is the reference case against which the other cases are compared.}
\label{fig:contributions}
\end{figure}

Fig.~\ref{fig:contributions} shows $\R_t$ and $\R_d$ (light and dark grey respectively) for each layout and different $W$ for $D$=3.38.
For case 0 the contribution from $\R_t$ increases from 7\% at $W$=0.13 to 13\% at $W$=0.26 and 0.39. It decays to 6\% for $W$=0.52. In the oscillating case studied by WR14, $\R_t$ scales linearly with the disc boundary layer thickness $\delta$, defined in RH13 and WR14 as a measure of the viscous diffusion from the disc surface. Using data from RH13, $\R_t$ also scales linearly with $\delta$ for steady rotation. Furthermore, $\R_t$ scales with coverage for $W$=0.13 for all layouts. The contribution to the overall drag reduction from $\R_d$ is negligible for cases 1 and 2 at all $W$, for which there is no spanwise interaction between the discs, and for all cases at $W$=0.13. The impact of the interdisc structures on drag reduction, synthesized by $\R_d$, becomes important for cases 3 and 4, whose $\R_t$ and $\R_d$ values are the same for the same $W$.

The cases for which $\R_d$ attains a finite value are boxed by the dashed line. Spanwise interaction between the discs must therefore be important for the formation of these structures, although at this stage it is still not clear why cases 3 and 4 have the same $\R_t$ and $\R_d$ values despite the shift of columns. For the cases boxed by the solid line, coverage scaling applies and structures do not appear, although in RH13 for $W$=0.26 and 0.39 the structures do contribute to the overall drag reduction.

\subsection{\label{sec:flowvis}Flow visualizations}
\begin{figure}
\centering
\includegraphics[width=0.9\textwidth]{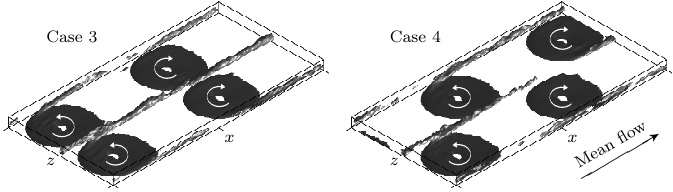}
\caption{Isosurfaces of $q$=$\sqrt{u_d^2+v_d^2+w_d^2}$=0.08 for case 3 (left) and case 4 (right), and $D$=3.38, $W$=0.52.}
\label{flowvis:case2}
\end{figure}

The contribution of $\R_d$ in cases 3 and 4 is proved to be important through the use of the FIK identity. Therefore, we resort to flow visualizations to display the interdisc structures that are responsible for $\R_d$.  Isosurfaces of $q$=$\sqrt{u_d^2+v_d^2+w_d^2}$=0.08 are shown in Fig.~\ref{flowvis:case2} for cases 3 and 4, the white arrows indicating the direction of disc rotation. In both cases the disc boundary layers are clearly visible.  The plots show the presence of the tubular structures first shown in RH13, elongated in the streamwise direction and situated between adjacent discs in the spanwise direction. For cases 1 and 2, the structures are instead not evident for similar values of $q$. The only instances where the structures are clearly visible occurs when there is spanwise interaction between the discs. This happens only for $W\hspace{-0.1cm}\geq0.26$ and for cases 0, 3, and 4, where the distance between the nearest disc centres is smaller than or equal to $\sqrt{2}D_0$.

\begin{figure}
\centering
\includegraphics[width=0.42\textwidth]{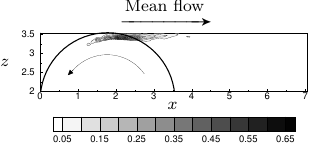}
\caption{Contour of $u_dv_d$ at $y^+$=14 for $D$=3.38, $W$=0.52, and case 3.}
\label{fig:q_bands_c2}
\end{figure}

A contour of $u_dv_d$ for case 3 at $y^+$=14 is shown in Fig.~\ref{fig:q_bands_c2}, indicating the disc side where the structure is created. The contour for case 4 is nearly identical. Differently from the experimental study by \citet{klewicki-hill-2003} of the laminar flow over a finite rotating surface patch, structures are not visible over both sides of the disc.  They do however propagate downstream parallel to the mean flow as the structures observed by \citeauthor{klewicki-hill-2003}.  Fig.~\ref{fig:q_bands_c2} shows that in all cases where there is a contribution from $\R_d$, the structures originate from the disc side where the wall forcing is along the upstream direction. When only one disc is included in the domain, the structures do not appear. Therefore the structures are created: i) when there is sufficient spanwise interaction between discs, i.e. $W\hspace{-0.1cm}\geq0.26$ and the distance between disc centres located in adjacent columns is smaller than or equal to $\sqrt{2}D_0$, and ii) at the disc sides where the wall streamwise motion is in the opposite direction to the mean flow.

\subsection{\label{sec:radial}Radial streaming}
\begin{figure}
\centering
\includegraphics[width=0.425\textwidth]{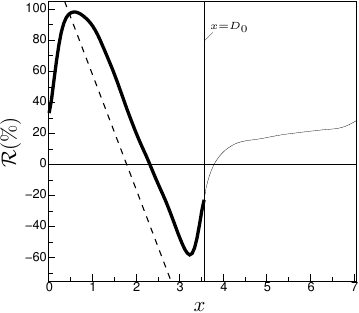}
\caption{Streamwise development of $\R$ along the disc centreline for case 3 for $W$$=$$0.52$. The thick line indicates the profile for the flow over the disc surface and the dashed line represents the drag reduction predicted by the laminar solution \eqref{eq:cf-lam}.}
\label{fig:radial-cf-c2}
\end{figure}

The FIK identity and flow visualizations of the structures have been useful to shed further light on the formation of the interdisc structures, but have not helped to explain the extra drag reduction effect with respect to coverage, discussed in Sec.~\ref{sec:layout}.  To gain more insight, since streamwise fixed-wall space is a common feature of the cases which present the additional drag reduction, the flow between discs is studied. The streamwise development of $\R$ along the disc centreline in case 3 is shown in Fig.~\ref{fig:radial-cf-c2} by the solid line. The drag reduction is non-zero at the disc centre and asymmetric about this point. A local peak of maximum drag reduction of 95\% occurs in the upstream disc region and intense drag increase appears in the downstream disc region.  Between discs there is a region of about $\R\hspace{-0.1cm}=$20\% that is responsible for the additional drag reduction with respect to coverage. This region must be created through the interaction between the mean flow and the disc flow because the net disc-flow wall-shear stress would be null if $u_m$=0, i.e. if the streamwise pressure gradient were absent, owing to the disc-flow symmetry.

By use of the laminar solution, the skin-friction coefficient is predicted as follows:
\begin{equation*}
C_{f,l}(x)=\frac{2}{U_b^2\Rp}
\left[
u_m^{\prime}(0) +
F_k
\left(\frac{2W}{D}\right)^{3/2}
\Rp^{1/2}x
\right]\text{,}
\numberthis
\label{eq:cf-lam}
\end{equation*}
where $F_k$=0.51 is given in \citet{schlichting-1979}. This prediction is not rigorous as the interaction between the mean and disc flow is not considered and end effects are neglected.  Despite this, as shown in Fig.~\ref{fig:radial-cf-c2}, the gradient of $\R$ with respect to $x$ is well predicted on the disc surface, although the drag reduction computed via the laminar solution is higher than $100\%$ due to flow reversal as the disc edge is not modelled. The DNS trend of $\R$ is shifted along $x$ by about 45$\nu^*/u_\tau^*$ relatively to the laminar prediction.  This is consistent with the streamwise shift in the disc flow of about 100$\nu^*/u_\tau^*$ observed at $y^+$=8 in the oscillating-disc case by WR14. This shift must be due to the interaction between the mean and disc flows, which is not considered in the laminar analysis.

\begin{figure}
\includegraphics[height=7cm]{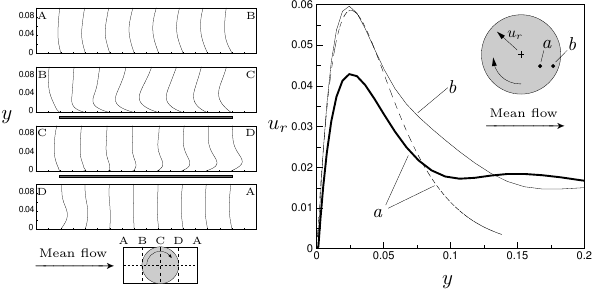}
\caption{Left: Streamwise disc flow $u_d$ vs. $y$ at different locations along the disc centreline for $W$=0.52. The letters A-D indicate which wall section the plot corresponds to. The plots above the grey bars correspond to locations on the disc surface. Right: Radial flow $u_r$ vs. $y$ for different locations on the disc surface. The solid lines represent turbulent profiles at locations {\emph a} (thick line) and {\emph b} (thin line), separated by 100$\nu^*/u_\tau^*$. The dashed line denotes the laminar profile at the {\emph a} location.}
\label{fig:radial-x}
\end{figure}

To further investigate the flow above the fixed-wall region between discs, the downstream development of $u_d$ along the centreline of the discs, shown in Fig.~\ref{fig:radial-x} (left), is studied.  The profiles are separated by 40$\nu^*/u_\tau^*$ and those on the disc surface are indicated by the grey bars.  From the beginning of the domain and up to about the disc centre, the disc creates a radial flow along the negative $x$ direction which retards the streamwise flow, thereby causing drag reduction. From the centre of the disc and up to the downstream disc tip, the radial flow enhances the streamwise  flow, resulting in drag increase. The radial flow is most energetic near the disc tips and this is represented by the peaks of drag reduction and drag increase in Fig.~\ref{fig:radial-cf-c2}.  The streamwise shift in the disc flow is also evident in Fig.~\ref{fig:radial-x} (left), shown by the switch from negative to positive $u_d$ occurring between points C and D at a distance of about 80$\nu^*/u_\tau^*$ downstream of the disc centre.

The disc flow persists further in the upstream direction than it does downstream, which explains the region of drag reduction above the fixed wall in Fig.~\ref{fig:radial-cf-c2}.  The disc flow upstream of a disc persists for 480$\nu^*/u_\tau^*$ from the upstream disc tip (point B), whereas the disc flow along the positive $x$ direction vanishes within a distance of only 120$\nu^*/u_\tau^*$ downstream of the disc tip (point D). In Fig.~\ref{fig:radial-x} (left) the peak of the $u_d$ profile varies above the disc, whereas in the laminar solution this location is invariant. The difference must be accounted for by the interaction of the disc flow with the mean streamwise flow.  Immediately off the disc surface the peak $y$-location of the disc flow increases by $\Delta y\hspace{-0.15cm}=$0.015.  As the wallward flow above the disc caused by the von K{\'a}rm{\'a}n pumping effect does not occur above the fixed wall, the radial flow is allowed to diffuse further into the channel.

Fig.~\ref{fig:radial-x} (right) presents the radial flow $u_r$ as a function of $y$ for two locations on the disc surface.  A graphical definition of $u_r$ is provided in Fig.~\ref{fig:radial-x} (inset).  The thick solid line is the radial flow above the disc at $x$=2.72, $z$=1.36, displaced by $r$=1.04 from the disc centre. The dashed line is the laminar prediction for the disc flow at the same $r$.  It is evident that at the same location the laminar and turbulent flow profiles do not coincide.  The thin solid line indicates the turbulent disc flow at a location 100$\nu^*/u_\tau^*$ downstream of the laminar prediction ($x$=3.27, $z$=1.36). At this location the turbulent and laminar profiles are almost identical for $y$\textless0.05, confirming the downstream shift of the disc flow.

\subsection{\label{sec:hd}Half-disc actuators}
\begin{figure}
\centering
\subfloat{\includegraphics[width=0.425\textwidth]{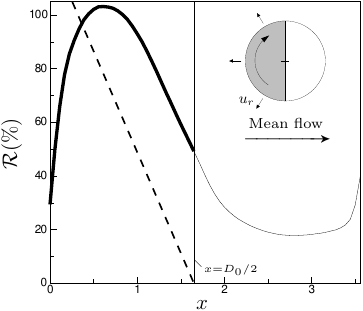}}\qquad
\subfloat{\raisebox{3.5cm}{
    \begin{tabular}{c|cc}
	\multicolumn{3}{c}{$D$$=$3.38} \\
	$W$	&    $\R_0(\%)$	&   $\R_h(\%)$	\\ \hline
	0.26	&	16.2	&	13.7	\\
	0.39	&	18.3	&	21.8	\\
	0.52	&	11.9	&	25.4	\\
	\multicolumn{1}{c}{}	&    \multicolumn{2}{c}{} \\
	\multicolumn{3}{c}{$D$$=$5.07} \\
 	$W$	&    $\R_0(\%)$	&   $\R_h(\%)$	\\ \hline
	0.26	&	17.5	&	13.0	\\
 	0.39	&	22.3	&	21.1	\\
	0.52	&	18.5	&	25.6
    \end{tabular}
}}
\caption{Left: Streamwise development of $\R$ along the half-disc centreline for $D$=3.38, $W$=0.52. The thick line indicates the profiles for the flow over the actuated half disc surface. The dashed line indicates the drag reduction predicted from the laminar solution \eqref{eq:cf-lam}. Inset: Schematic of a half-disc actuator. Right: Performance data for half-disc simulations.}
\label{fig:radial-cf-hd}
\end{figure}

As evidenced by Fig.~\ref{fig:radial-cf-c2} the radial flow induced by the downstream half of the discs causes drag increase.  To eliminate this effect, a half-disc configuration is studied, whereby the downstream disc half is covered and the wall-velocity is zero. The half-disc actuators are investigated for $D$=3.38,5.07 and $W$=0.13,0.26,0.39.  The drag reduction data for the half-disc simulations (subscript $h$) are presented in the table in Fig.~\ref{fig:radial-cf-hd} (right) with the corresponding data for case 0 (subscript $0$). As shown in Fig.~\ref{fig:radial-cf-hd} (left), the negative effect of the downstream radial flow is eliminated by covering this portion of the disc. The azimuthal flow, which contributes favourably to drag reduction, is also removed. As expected, the prediction of the laminar solution (dashed lines) is worse than in the full-disc case.

For both disc diameters and $W$=0.26, the drag reduction decreases when the downstream disc half is covered. This is because for low $W$ the negative effect of the radial flow is less important than the benefit of the azimuthal forcing.  For $W$\textgreater0.26 the drag reduction increases when the downstream disc half is covered and a maximum $\R_h$=25.6\% is computed.  For high $W$ the removal of the downstream disc section and the associated radial flow therefore outweighs the loss of beneficial effects induced by the azimuthal flow.

Although the increased drag reduction from this configuration is an interesting result, our model contains many simplifications. In an experimental set up a step would occur between the covered and uncovered halves of the disc, resulting in recirculation regions.  Neither this nor any interaction between the mean flow and the disc housing is considered.  A novel flow-control device has been realized experimentally by \citet{koch-kozulovic-2013} who performed boundary layer experiments on a disc set up with one spanwise half covered.  Differently from our actuators this is a passive method as the disc motion is driven by the mean flow and there is no external power input.  As the uncovered disc half rotates, the velocity difference between the mean flow and the wall decreases, thereby reducing the wall-shear stress while drawing energy from the mean flow.

A discussion must be included on the categorization of flow control methods as either drag reduction or pumping \cite{hoepffner-fukagata-2009}. For the original disc actuators, studied by RH13 (case 0 in Fig. \ref{fig:rh13discs}), although a mean flow is induced by the discs in the absence of streamwise pressure gradient, this mean flow is null when averaged along the streamwise direction. Therefore RH13's disc-flow control method can be categorized as drag reduction. For the half-disc technique, a net upstream mean flow is instead created in the absence of streamwise pressure gradient as an indirect response to the wall forcing, whose average in either the spanwise or streamwise direction is null. The half-disc method can thus be classified as indirect pumping. Direct pumping would instead occur if the reduction of wall friction were induced by a body force or a wall velocity distribution which are not zero when averaged along the streamwise direction.

\subsection{\label{sec:ann}Annular actuators}
\begin{figure}
  \includegraphics[width=0.85\textwidth]{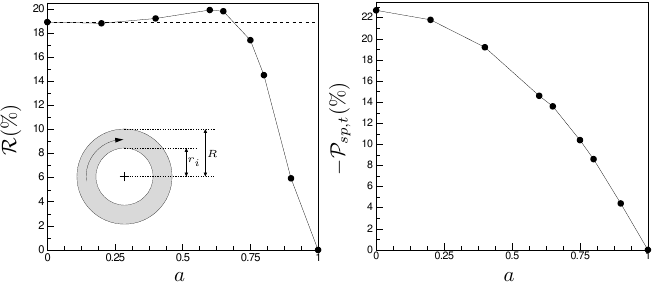}
  \caption{Performance quantities vs. annulus ratio, $a$=$r_i/R$, for $D$=3.38 and $W$=0.39. Left: Drag reduction, $\R$.  Right: Power spent, $\Pt$.  Inset: Schematic of an annular actuator.}
  \label{fig:annular-actuator}
\end{figure}

The laminar solution provides further direction for improvement of the disc-flow technique. The wallward flow produced by the von K{\'a}rm{\'a}n pump, which is uniform over the disc surface in planes parallel to the wall, can be expected to direct the streamwise flow towards the wall, causing a detrimental effect to drag reduction. Furthermore, the azimuthal forcing near the disc centre is of low velocity and, as shown in Sec.~\ref{sec:ann-gap}, the large-scale forcing appears to be important for drag reduction.  Therefore, annular actuators are studied, with the intent of attenuating the wallward flow and eliminating the low velocity motion near the disc centre, which is thought to have a marginal contribution to drag reduction. The ratio of the internal and external radii, $a$=$r_i/R$, is varied from 0 to 1, and the drag reduction and power spent are shown as functions of $a$ in Fig.~\ref{fig:annular-actuator}.  A schematic of the actuators is shown in Fig.~\ref{fig:annular-actuator} (inset).

The drag reduction remains approximately constant at $\R$=19\% for $a$\textless0.375. An optimum of $\R$=20\% is reached at $a$=0.6, beyond which the drag reduction decreases.  This confirms the prediction that the flow induced near the disc centre has an overall negative effect on drag reduction.  Beyond the optimum $a$=0.6 the removal of the central part of the disc causes a sharp decrease in $\R$ to a null value for $a=1$.

The power spent, shown in Fig.~\ref{fig:annular-actuator}, instead shows a rapid monotonic decrease as $a$ increases. Analogously to the changes due to the gap size, shown in Fig.~\ref{fig:gaptable}, the response of the power spent to the change in wall boundary conditions is more significant than for the drag reduction.

\subsection{\label{sec:spect}Spectral truncation}
\begin{figure}
\centering
\includegraphics[width=0.85\textwidth]{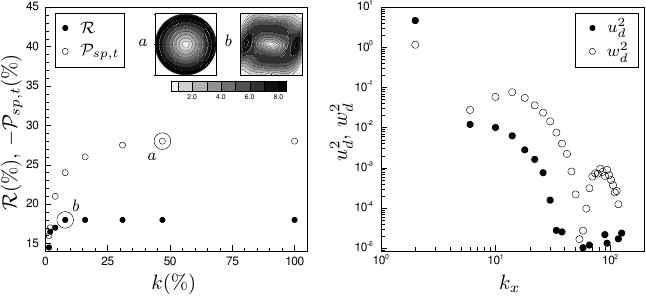}
\caption{Left: Effect of spectral truncation on performance quantities. $\R$ and $\Pt$ vs. $k$, the proportion of modes synthesizing the wall boundary conditions. Inset: Contours of $q$=$\sqrt{u_d^2+w_d^2}$ at $y$=0 for the circled cases for $k$=47\% (left) and $k$=8\% (right). Right: Energy of streamwise and spanwise forcing, measured by $u_d^2$ and $w_d^2$, vs. the streamwise wavenumber $k_x$.}
\label{fig:spect}
\end{figure}

The investigation of annular actuators confirms that the large scale forcing is important for drag reduction. The spectral representation of the boundary conditions is therefore examined to elucidate the effects of large and small scale forcing.  By truncating the number of Fourier modes that describe the disc motion, it is possible to force only a specified range of scales.  The proportion of modes forced in the homogeneous directions is given by $k(\%)$=100$k_{f,i}/N_i$, where $k_{f,i}$ is the maximum forced wavenumber, $N_i$ is the total number of modes, and the $i$ subscript denotes the streamwise or spanwise direction.  The truncation of modes is symmetrical in each direction, and so $k$=100$k_x/N_x$=100$k_z/N_z$.  The drag reduction and power spent are plotted as functions of $k$ in Fig.~\ref{fig:spect} (left). As the number of forced modes increases, both $\R$ and $\Pt$ asymptotically approach the values given when all of the modes are included. The drag reduction reaches the asymptotic value only when $k$=8\%, while $\Pt$ reaches the asymptote when $k$=47\%.  The contour plots of azimuthal wall velocity for these truncations are shown in Fig.~\ref{fig:spect} (insets). Fig.~\ref{fig:spect} (right) displays the energy contained within the streamwise modes. A large proportion of the energy is contained within the low wavenumber modes.  The energy of the wall streamwise velocity has a peak value at $k_x$=2, then drops monotonically with $k_x$ up to about $k_x=50$, at which it attains small values comprised between 10$^{-5}$ and 10$^{-6}$.  The energy of the wall spanwise velocity has peaks of amplitude decreasing continuously by more than one order of magnitude and occurring at $k_x$=2, 14 and 82.  These peaks are separated by minima at $k_x$=6 and 54 of magnitude 10$^{-2}$ and 10$^{-5}$, respectively.

The results in Fig.~\ref{fig:spect} (left) bear analogy with the effects of gap size and annular actuators on the performance quantities, presented in Sec.~\ref{sec:ann-gap} and \ref{sec:ann}, respectively.  In all cases it is evident that the large scale forcing is most responsible for the drag reduction, shown by the lack of significant change in $\R$ when high-wavenumber modes are eliminated from the disc spectral representation, the gap size is increased, or the central part of the disc is removed.  This is significant as it means that low-order models, which only capture prescribed features of the turbulence dynamics, might be sufficient for computing accurate values of drag reduction. The boundary conditions have also been modified to only force either the spanwise or streamwise wall velocity. Drag increase occurred in both cases.  This shows that a fully nonlinear mechanism must be responsible for drag reduction.

\section{\label{sec:summary}Summary}
This paper has presented results on the rotating disc method for drag reduction. A summary of these results is presented herein.
\begin{itemize*}
\item The effects of coverage and layout on the performance of the disc technique have been investigated, with unexpected gains in $\R$ found upon the removal of discs.  For example for disc-tip velocity $W$=0.52 the removal of half of the discs leads to an increase in $\R$. At this $W$, an optimal spacing of 1.5$D_0$ between disc centres results in an additional drag reduction of $\R$=5\% relatively to the RH13 layout.  For intermediate values of $W$, the gain in $\R$ always occurs when streamwise space of stationary wall occurs between discs.
\item For low $W$, the drag reduction scales linearly with coverage and is well predicted from the standing-wave data by \citet{viotti-quadrio-luchini-2009} through scaling factors to account for the changes in waveform, angle of wall forcing, and coverage.
\item The power spent to actuate the discs is well predicted by the laminar solution, does not depend on the disc arrangement, and scales with coverage for all $W$.
\item The FIK identity and flow visualizations have been useful to elucidate the criteria for the formation of structures appearing between discs.  The structures are created only when there is sufficient interaction between spanwise neighbouring discs and at the disc sides where the wall streamwise motion is in the opposite direction to the mean flow. The disc-tip velocity must be $W\hspace{-0.15cm}\geq$0.26 and the maximum spacing between disc centres in neighbouring columns must be $\sqrt{2}D_0$, where $D_0$ is the outer diameter of the circle occupied by the disc and the annular gap.
\item It has been shown that the radial flow due to the von K{\'a}rm{\'a}n pumping effect creates a viscous layer over areas of stationary wall between discs. This boundary layer is responsible the the additional drag reduction with respect to the value predicted through the scaling with the actuated area.
\item Novel half-disc and annular actuators have been simulated to improve the drag reduction effect, resulting in a maximum of $\R$=26\%. A comparison between these disc-flow drag reduction data and those of other drag reduction techniques is given in Table~\ref{tab:performance-comparison}.
\begin{table*}[h!]
\centering
    \begin{tabular}{l|c|l}
Control strategy & $ \R_{max}(\%)$ & Details \\ \hline
Riblets \citep{sasamori-etal-2014} & 12 & Sinusoidal riblets with spanwise modulation \\
Opposition $v$-control \citep{choi-moin-kim-1994} & 25 & Control with wall-normal velocity \\
Opposition $w$-control \citep{choi-moin-kim-1994} & 30 & Control with spanwise velocity \\
Oscillating wall \citep{quadrio-ricco-2004} & 45 & Oscillation period, $T^+$=100.  Amplitude, $W^+$=27 \\ \hline
Steadily rotating discs (RH13) & 23 & $D^+$=801, $W^+$=10.2  \\
Oscillating discs (WR14) & 20 & $D^+$=812, $W^+$=13.5, $T^+$=794 \\ 
Annular actuators & 20 &  $D^+$=514, $W^+$=10.1, $a$=0.6 \\
Half-disc actuators & 26 & $D^+$=743, $W^+$=14.0
   \end{tabular}
\caption{Comparison of the disc-flow drag reduction data with the ones from other control strategies. Larger drag reduction values may be found for the annular and half-disc actuators as a full optimization has not been performed.}
\label{tab:performance-comparison}
\end{table*}
\item According to the categorization proposed by \citet{hoepffner-fukagata-2009}, the original disc actuators studied by RH13 have been classified as a drag reduction method. The half-disc actuators have instead been classified as an indirect pumping method. The term pumping arises from the net upstream flow that would be created by the half discs even without streamwise pressure gradient, while the term indirect indicates that this upstream flow is engendered even though the forcing at the wall is null when averaged along the streamwise direction.
\item The effect of the forcing scales on the drag reduction and on the power spent has also been studied. Truncation of the number of forced modes in the boundary conditions has shown that it is the larger scales that most contribute to drag reduction. The power spent has a more marked dependence on the precise spectral representation of the wall forcing than drag reduction.
\end{itemize*}

\begin{acknowledgments}
We would like to thank the Department of Mechanical Engineering at the University of Sheffield for funding this research. This work would have not been possible without the use of the computing facilities of N8 HPC, funded by the N8 consortium and EPSRC (Grant EP/K000225/1) and coordinated by the Universities of Leeds and Manchester.  Our thanks also go to Prof. J.F. Morrison for recommending the \citeauthor{klewicki-hill-2003} paper and to Mr Alessandro Melis for his insightful comments on a preliminary version of the manuscript.
\end{acknowledgments}

\appendix
\section{\label{app:drresults}Table of drag reduction data}
\noindent The data for $\R$ and $\Pt$ are given in Table~\ref{tab:results}.
\begin{table*}[h!]
\centering
    \begin{tabular}{ccc|cc}
								& 	Case	&	$W$	&    	$\R(\%)$&	$-\Pt(\%)$ \\ \hline
\multirow{4}{*}{\scalebox{0.85}{$\Box$}} 			&	0	&	0.13	&	7.2	&	2.4	\\
								&	0	&	0.26	&	16.2	&	9.8	\\
								&	0	&	0.39	&	18.3	&	22.7	\\
								&	0	&	0.52	&	11.9	&	42.5	\\ \hline
\multirow{4}{*}{$\triangledown$}				&	1	&	0.13	&	1.7 	&	0.6	\\
								&	1	&	0.26	&	5.1	&	2.5	\\
								&	1	&	0.39	&	7.4	&	5.8	\\
								&	1	&	0.52	&	8.5	&	10.9	\\ \hline
\multirow{4}{*}{$\vartriangle$}					&	2	&	0.13	&	3.5	&	1.2	\\
								&	2	&	0.26	&	8.2	&	5.0	\\
								&	2	&	0.39	&	9.4	&	11.4	\\
								&	2	&	0.52	&	8	&	21.4
    \end{tabular}
    \quad
    \begin{tabular}{ccc|cc}
								& 	Case	&	$W$	&    	$\R(\%)$&	$-\Pt(\%)$ \\ \hline
\multirow{4}{*}{\scalebox{1.6}{$\circ$}}			&	3	&	0.13	&	3.6	&	1.2	\\
								&	3	&	0.26	&	10.3	&	5.0	\\
								&	3	&	0.39	&	14.0	&	11.6	\\
								&	3	&	0.52	&	15.5	&	21.7	\\ \hline
\multirow{4}{*}{\scalebox{0.79}{\rotatebox{45}{$\Box$}}}	&	4	&	0.13	&	3.6	&	1.2	\\
								&	4	&	0.26	&	10.3	&	5.0	\\
								&	4	&	0.39	&	14.0	&	11.6	\\
								&	4	&	0.52	&	15.5	&	21.6	\\ \hline
\multirow{4}{*}{$\rhd$}						&	5	&	0.13	&	8.0	&	2.7	\\
								&	5	&	0.26	&	17.0	&	9.7	\\
								&	5	&	0.39	&	18.5	&	24.7	\\
								&	5	&	0.52	&	14.8	&	46.4
    \end{tabular}
\caption{Performance data for different forcing conditions and layouts.}
\label{tab:results}
\end{table*}

\section{\label{app:psp}Laminar power spent calculations}
The laminar flow solutions to the flows induced by spinning and oscillating discs were used by RH13 and WR14 to predict the work done to enforce the disc motion. Therein the laminar power spent to actuate the discs is calculated as the ratio between the power spent to actuate the discs, $\Pl$, and the power spent to drive the fluid in the streamwise direction, $\Px$.  The efficiency of the mechanical system used to power the discs is not considered in the computation of either $\Pl$ or $\Pt$.  RH13 and WR14 considered $\Pl$ as being the volume-averaged power spent above the disc surface (i.e. averaged over $\pi D^2h/4$).  This is equivalent to computing the power spent averaged over the actuated wall area.  $\Px$ was computed as the average over the volume $D_0^2h$. The contribution to the power spent due to the annular flow between the disc and the stationary wall was not considered. In the following $\Pl$ is averaged over the whole wetted area for a meaningful comparison with the power spent computed through DNS. The contribution of the gap flow to the power spent is also accounted for. The derivations of the adjusted
formulae are outlined below.

\begin{figure}
  \centering
  \includegraphics[width=0.8\textwidth]{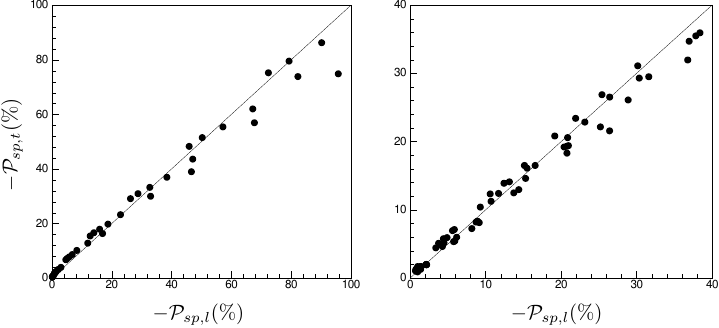}
  \caption{Left: $\Pt$ vs. $\Pl$ for data from RH13. $\Pl$ is computed through \eqref{eq:psp-rh13}. Right: $\Pt$ vs. $\Pl$, for data from WR14. $\Pl$ is computed through \eqref{eq:plam-osc}.}
\label{pspent-rh13}
\end{figure}

By taking the volume integral of the viscous stresses work term in equation (1-108) of \citet{hinze-1975} as follows
\begin{equation*}
\mathcal{P}_{sp,l}=\frac{\nu^*}{L_x^*L_y^*L_z^*}\int_0^{L_x^*}\int_0^{L_y^*}\int_0^{L_z^*}
\frac{\p}{\p x_i^*}
\left[
u_j^*
\left(
\frac{\p u_i^*}{\p x_j^*}+\frac{\p u_j^*}{\p x_i^*}
\right)
\right]
\mathrm{d}x^*\mathrm{d}y^*\mathrm{d}z^*\text{,}
\numberthis
\label{eq:vis-stress}
\end{equation*}
the work done by the viscous stresses per unit time is obtained. The Einstein summation of repeated indices is used in \eqref{eq:vis-stress}.  The decomposition of the flow field given in \eqref{eq:decomp} is used and only ${\bf u}_d$ is retained as neither a mean streamwise flow nor any turbulent fluctuations are taken into account. Substituting $u_d$=$u_\theta\cos\theta$ and $w_d$=$u_\theta\sin\theta$ in \eqref{eq:vis-stress}, and changing to cylindrical coordinates leads to
\begin{equation}
\mathcal{P}_{sp,l}=\frac{\nu^*}{D_0^{*2}}\int_0^{L_y^*}\int_0^{D_0^*/2}\int_0^{2\pi}
u_\theta^*\frac{\p u_\theta^*}{\p y^*}
r^*\mathrm{d}\theta\mathrm{d}r^*\mathrm{d}y^*\text{.}
\label{eq:pd-int}
\end{equation}
There are two distinct intervals over which the integral must be taken. The first considers the disc surface (i.e. for $r$$\leq$$D^*/2$, $u_\theta^*$=2$WG(\eta)r^*/D^*$, where $G(\eta)$ is tabulated by \citet{schlichting-1979} and $\eta=y^*\sqrt{2W^*/(\nu^* D^*)}$ is the scaled wall-normal coordinate) and the second considers the annular flow for $D^*/2$$<$$r^*$$<$$D_0^*/2$. To include the gap into the calculation it is assumed that within this region the wall-normal scaling remains the same as the von K{\'a}rm{\'a}n solution and that the angular velocity within this region is therefore given by $u^*_{\theta,g}$=$W^*G(\eta)(D_0^*/2-r^*)/c^*$. Expression \eqref{eq:pd-int} then becomes
\begin{equation*}
 \mathcal{P}_{sp,l}=\frac{\nu^*}{D_0^{*2}}\left(
\int_0^{L_y^*}\int_0^{D^*/2}\int_0^{2\pi}
u_\theta^*\frac{\p u_\theta^*}{\p y^*}
r^*\mathrm{d}\theta\mathrm{d}r^*\mathrm{d}y^*
+
\int_0^{L_y^*}\int_{D^*/2}^{D_0^*/2}\int_0^{2\pi}
u_{\theta,g}^*\frac{\p u_{\theta,g}^*}{\p y^*}
r^*\mathrm{d}\theta\mathrm{d}r^*\mathrm{d}y^*
\right)\text{.}
\end{equation*}
Upon substituting the definitions of $u_\theta$ and $u_{\theta,g}$ and integrating, one finds
\begin{equation*}
\mathcal{P}_{sp,l}=\frac{\pi G_k W^{*5/2}}{D_0^{*2}} \sqrt{\frac{2\nu^*}{D^*}}
\left(
\frac{D^{*2}}{8}+\frac{c^*D^*}{3}+\frac{c^{*2}}{6}
\right)\text{.}
\numberthis
\label{eq:psp-d}
\end{equation*}
Dividing \eqref{eq:psp-d} by the power spent to drive the fluid in the streamwise direction and scaling in outer units yields the formula for the percent laminar power spent to move the discs,
\begin{equation*}
\Pl(\%)=
\frac{100\pi G_k \Rp^{3/2} W^{5/2}}{U_b\Rt^2 D_0^2}
\sqrt{\frac{2}{D}}
\left(
\frac{D^2}{8}+\frac{cD}{3}+\frac{c^2}{6}
\right)\text{,}
\numberthis
\label{eq:psp-rh13}
\end{equation*}
Fig.~\ref{pspent-rh13} (left) presents the RH13 data for $\Pt$ versus $\Pl$ computed from formula \eqref{eq:psp-rh13}. The agreement of $\Pl$ with the DNS data is much better with the corrected averaging and improvement.

The laminar power spent formulae presented in WR14 are now derived to incorporate the annular clearance flow. Formula (3.6) in WR14 is amended and improved as follows
\begin{align*}
\Pl^*=\frac{\pi^{3/2}\mathcal{G}(\gamma)W^{*2}}{D_0^{*2}}\sqrt{\frac{\nu^*}{T^*}}\left(\frac{D^{*2}}{8}+\frac{c^*D^*}{3}+\frac{c^{*2}}{6}\right)\text{,}
\end{align*}
where $\mathcal{G}(\gamma)$$=$$(2\pi)^{-1}\int_0^{2\pi}G(0,t)G'(0,t)\textup{d}t$ and $\gamma$=$T^*W^*/(\pi D^*)$. Dividing by $\mathcal{P}_x^*$ and scaling in outer units yields $\Pl$ expressed as a percentage of the power spent to drive the fluid in the $x$ direction,
\begin{align*}
\Pl(\%)=\frac{100 (\pi \Rp)^{3/2} \mathcal{G}(\gamma) W^2}{U_b \Rt^2 D_0^2 \sqrt{T}}\left( \frac{D^2}{8}+\frac{cD}{3}+\frac{c^2}{6} \right)\text{,}
\numberthis
\label{eq:plam-osc}
\end{align*}
which is amended formula (3.8) in WR14. Fig.~\ref{pspent-rh13} (right) shows a much improved agreement of $\Pl$ with the DNS data for the oscillating-disc flow as well. An analytical approximation to $\mathcal{G}$ for $\gamma$$\ll$1 is given in equation (3.10) of WR14.  In the limit $\gamma$$\ll$1 \citet{rosenblat-1959} derives a first-order approximation to $u_\theta^*$.  Upon substituting this approximation into \eqref{eq:vis-stress} and integrating the viscous stresses over the volume, the first-order approximation to $\Pl^*$ is found. Expressed as a percentage of $\mathcal{P}_x$, this is
\begin{equation*}
\mathcal{P}_{sp,l,\gamma\ll1}(\%)=\frac{-50(\pi\Rp)^{3/2} W^2}{U_b \Rt^2 D_0^2 \sqrt{T}}\left( \frac{D^2}{8}+\frac{cD}{3}+\frac{c^2}{6} \right)\text{.}
\end{equation*}
The asymptotic limit of $\mathcal{G}$ for $\gamma$$\gg$1 is found by WR14 to be $\mathcal{G}_{\gamma\gg1}$$=$$G_s\sqrt{\gamma/2}$, where $G_s$$=$$-$0.61592 is given in \citet{rogers-lance-1960}. By substituting this into \eqref{eq:plam-osc} the asymptotic form of the power spent in the limit $\gamma$$\gg$1 is found
\begin{equation*}
\mathcal{P}_{sp,l,\gamma\gg1}(\%)=\frac{100 \pi G_s \Rp^{3/2}W^{5/2}}{U_b \Rt^2 D_0^2 \sqrt{2D}}\left( \frac{D^2}{8}+\frac{cD}{3}+\frac{c^2}{6} \right)\text{.}
\end{equation*}

We close this appendix with a note on the power transfer to and from the discs. The spatial distribution of the power spent is presented in Fig.~11 of RH13.  Therein it is stated that the areas for which this power is positive indicate regions where the fluid performs work on the disc, and that this is a spatially localized regenerative braking effect.  This latter terminology is used incorrectly, as pointed out by Prof. J.F. Morrison (personal communication).  Although it is true that over these areas the disc motion is aided by the fluid, no energy can be extracted or stored. For this reason the term `regenerative braking' does not apply to the steadily rotating discs.  For the oscillating wall however the phenomenon occurs in time.  Therefore as for some phases of the oscillation the net power transfer to the wall is positive over the whole wetted area, it could theoretically be possible for the energy to be stored and reused. In this instance, the term regenerative braking is
appropriate.

\maketitle

\bibliography{pr}

\begin{thebibliography}{38}%
\makeatletter
\providecommand \@ifxundefined [1]{%
 \@ifx{#1\undefined}
}%
\providecommand \@ifnum [1]{%
 \ifnum #1\expandafter \@firstoftwo
 \else \expandafter \@secondoftwo
 \fi
}%
\providecommand \@ifx [1]{%
 \ifx #1\expandafter \@firstoftwo
 \else \expandafter \@secondoftwo
 \fi
}%
\providecommand \natexlab [1]{#1}%
\providecommand \enquote  [1]{``#1''}%
\providecommand \bibnamefont  [1]{#1}%
\providecommand \bibfnamefont [1]{#1}%
\providecommand \citenamefont [1]{#1}%
\providecommand \href@noop [0]{\@secondoftwo}%
\providecommand \href [0]{\begingroup \@sanitize@url \@href}%
\providecommand \@href[1]{\@@startlink{#1}\@@href}%
\providecommand \@@href[1]{\endgroup#1\@@endlink}%
\providecommand \@sanitize@url [0]{\catcode `\\12\catcode `\$12\catcode
  `\&12\catcode `\#12\catcode `\^12\catcode `\_12\catcode `\%12\relax}%
\providecommand \@@startlink[1]{}%
\providecommand \@@endlink[0]{}%
\providecommand \url  [0]{\begingroup\@sanitize@url \@url }%
\providecommand \@url [1]{\endgroup\@href {#1}{\urlprefix }}%
\providecommand \urlprefix  [0]{URL }%
\providecommand \Eprint [0]{\href }%
\providecommand \doibase [0]{http://dx.doi.org/}%
\providecommand \selectlanguage [0]{\@gobble}%
\providecommand \bibinfo  [0]{\@secondoftwo}%
\providecommand \bibfield  [0]{\@secondoftwo}%
\providecommand \translation [1]{[#1]}%
\providecommand \BibitemOpen [0]{}%
\providecommand \bibitemStop [0]{}%
\providecommand \bibitemNoStop [0]{.\EOS\space}%
\providecommand \EOS [0]{\spacefactor3000\relax}%
\providecommand \BibitemShut  [1]{\csname bibitem#1\endcsname}%
\let\auto@bib@innerbib\@empty
\bibitem [{\citenamefont {Gad-el Hak}(2000)}]{gadelhak-2000}%
  \BibitemOpen
  \bibfield  {author} {\bibinfo {author} {\bibfnamefont {M.}~\bibnamefont
  {Gad-el Hak}},\ }\href@noop {} {\emph {\bibinfo {title} {Flow control -
  {P}assive, {A}ctive and {R}eactive {F}low {M}anagement}}}\ (\bibinfo
  {publisher} {Cambridge University Press},\ \bibinfo {year}
  {2000})\BibitemShut {NoStop}%
\bibitem [{\citenamefont {Kasagi}\ \emph {et~al.}(2009)\citenamefont {Kasagi},
  \citenamefont {Suzuki},\ and\ \citenamefont
  {Fukagata}}]{kasagi-suzuki-fukagata-2009}%
  \BibitemOpen
  \bibfield  {author} {\bibinfo {author} {\bibfnamefont {N.}~\bibnamefont
  {Kasagi}}, \bibinfo {author} {\bibfnamefont {Y.}~\bibnamefont {Suzuki}}, \
  and\ \bibinfo {author} {\bibfnamefont {K.}~\bibnamefont {Fukagata}},\
  }\bibfield  {title} {\enquote {\bibinfo {title} {Micromechanical
  systems-based feedback control of turbulence for skin friction reduction},}\
  }\href@noop {} {\bibfield  {journal} {\bibinfo  {journal} {Ann. Rev. Fluid
  Mech.}\ }\textbf {\bibinfo {volume} {41}},\ \bibinfo {pages} {231} (\bibinfo
  {year} {2009})}\BibitemShut {NoStop}%
\bibitem [{\citenamefont {Wilkinson}(1990)}]{wilkinson-1990}%
  \BibitemOpen
  \bibfield  {author} {\bibinfo {author} {\bibfnamefont {S.~P.}\ \bibnamefont
  {Wilkinson}},\ }\enquote {\bibinfo {title} {Viscous {D}rag {R}eduction in
  {B}oundary {L}ayers},}\ \ (\bibinfo  {publisher} {AIAA},\ \bibinfo {year}
  {1990})\ Chap.\ \bibinfo {chapter} {Interactive wall turbulence control}, p.\
  \bibinfo {pages} {479}\BibitemShut {NoStop}%
\bibitem [{\citenamefont {Jung}\ \emph {et~al.}(1992)\citenamefont {Jung},
  \citenamefont {Mangiavacchi},\ and\ \citenamefont
  {Akhavan}}]{jung-mangiavacchi-akhavan-1992}%
  \BibitemOpen
  \bibfield  {author} {\bibinfo {author} {\bibfnamefont {W.J.}\ \bibnamefont
  {Jung}}, \bibinfo {author} {\bibfnamefont {N.}~\bibnamefont {Mangiavacchi}},
  \ and\ \bibinfo {author} {\bibfnamefont {R.}~\bibnamefont {Akhavan}},\
  }\bibfield  {title} {\enquote {\bibinfo {title} {Suppression of turbulence in
  wall-bounded flows by high-frequency spanwise oscillations},}\ }\href@noop {}
  {\bibfield  {journal} {\bibinfo  {journal} {Phys. Fluids A}\ }\textbf
  {\bibinfo {volume} {4}},\ \bibinfo {pages} {1605} (\bibinfo {year}
  {1992})}\BibitemShut {NoStop}%
\bibitem [{\citenamefont {Laadhari}\ \emph {et~al.}(1994)\citenamefont
  {Laadhari}, \citenamefont {Skandaji},\ and\ \citenamefont
  {Morel}}]{laadhari-skandaji-morel-1994}%
  \BibitemOpen
  \bibfield  {author} {\bibinfo {author} {\bibfnamefont {F.}~\bibnamefont
  {Laadhari}}, \bibinfo {author} {\bibfnamefont {L.}~\bibnamefont {Skandaji}},
  \ and\ \bibinfo {author} {\bibfnamefont {R.}~\bibnamefont {Morel}},\
  }\bibfield  {title} {\enquote {\bibinfo {title} {Turbulence reduction in a
  boundary layer by local spanwise oscillating surface},}\ }\href@noop {}
  {\bibfield  {journal} {\bibinfo  {journal} {Phys. Fluids}\ }\textbf {\bibinfo
  {volume} {6}},\ \bibinfo {pages} {3218} (\bibinfo {year} {1994})}\BibitemShut
  {NoStop}%
\bibitem [{\citenamefont {Viotti}\ \emph {et~al.}(2009)\citenamefont {Viotti},
  \citenamefont {Quadrio},\ and\ \citenamefont
  {Luchini}}]{viotti-quadrio-luchini-2009}%
  \BibitemOpen
  \bibfield  {author} {\bibinfo {author} {\bibfnamefont {C.}~\bibnamefont
  {Viotti}}, \bibinfo {author} {\bibfnamefont {M.}~\bibnamefont {Quadrio}}, \
  and\ \bibinfo {author} {\bibfnamefont {P.}~\bibnamefont {Luchini}},\
  }\bibfield  {title} {\enquote {\bibinfo {title} {Streamwise oscillation of
  spanwise velocity at the wall of a channel for turbulent drag reduction},}\
  }\href@noop {} {\bibfield  {journal} {\bibinfo  {journal} {Phys. Fluids}\
  }\textbf {\bibinfo {volume} {21}} (\bibinfo {year} {2009})}\BibitemShut
  {NoStop}%
\bibitem [{\citenamefont {Choi}\ \emph {et~al.}(2002)\citenamefont {Choi},
  \citenamefont {Xu},\ and\ \citenamefont {Sung}}]{choi-xu-sung-2002}%
  \BibitemOpen
  \bibfield  {author} {\bibinfo {author} {\bibfnamefont {J-I.}\ \bibnamefont
  {Choi}}, \bibinfo {author} {\bibfnamefont {C-X.}\ \bibnamefont {Xu}}, \ and\
  \bibinfo {author} {\bibfnamefont {H.J.}\ \bibnamefont {Sung}},\ }\bibfield
  {title} {\enquote {\bibinfo {title} {Drag reduction by spanwise wall
  oscillation in wall-bounded turbulent flows},}\ }\href@noop {} {\bibfield
  {journal} {\bibinfo  {journal} {AIAA J.}\ }\textbf {\bibinfo {volume} {40}},\
  \bibinfo {pages} {842} (\bibinfo {year} {2002})}\BibitemShut {NoStop}%
\bibitem [{\citenamefont {Skote}(2011)}]{skote-2011}%
  \BibitemOpen
  \bibfield  {author} {\bibinfo {author} {\bibfnamefont {M.}~\bibnamefont
  {Skote}},\ }\bibfield  {title} {\enquote {\bibinfo {title} {Turbulent
  boundary layer flow subject to streamwise oscillation of spanwise
  wall-velocity},}\ }\href@noop {} {\bibfield  {journal} {\bibinfo  {journal}
  {Phys. Fluids}\ }\textbf {\bibinfo {volume} {23}},\ \bibinfo {pages} {081703}
  (\bibinfo {year} {2011})}\BibitemShut {NoStop}%
\bibitem [{\citenamefont {Skote}(2013)}]{skote-2013}%
  \BibitemOpen
  \bibfield  {author} {\bibinfo {author} {\bibfnamefont {M.}~\bibnamefont
  {Skote}},\ }\bibfield  {title} {\enquote {\bibinfo {title} {Comparison
  between spatial and temporal wall oscillations in turbulent boundary layer
  flows},}\ }\href@noop {} {\bibfield  {journal} {\bibinfo  {journal} {J. Fluid
  Mech.}\ }\textbf {\bibinfo {volume} {730}},\ \bibinfo {pages} {273} (\bibinfo
  {year} {2013})}\BibitemShut {NoStop}%
\bibitem [{\citenamefont {Quadrio}\ \emph {et~al.}(2009)\citenamefont
  {Quadrio}, \citenamefont {Ricco},\ and\ \citenamefont
  {Viotti}}]{quadrio-ricco-viotti-2009}%
  \BibitemOpen
  \bibfield  {author} {\bibinfo {author} {\bibfnamefont {M.}~\bibnamefont
  {Quadrio}}, \bibinfo {author} {\bibfnamefont {P.}~\bibnamefont {Ricco}}, \
  and\ \bibinfo {author} {\bibfnamefont {C.}~\bibnamefont {Viotti}},\
  }\bibfield  {title} {\enquote {\bibinfo {title} {Streamwise-travelling waves
  of spanwise wall velocity for turbulent drag reduction},}\ }\href@noop {}
  {\bibfield  {journal} {\bibinfo  {journal} {J. Fluid Mech.}\ }\textbf
  {\bibinfo {volume} {627}},\ \bibinfo {pages} {161} (\bibinfo {year}
  {2009})}\BibitemShut {NoStop}%
\bibitem [{\citenamefont {Gouder}\ \emph {et~al.}(2013)\citenamefont {Gouder},
  \citenamefont {Potter},\ and\ \citenamefont
  {Morrison}}]{gouder-potter-morrison-2013}%
  \BibitemOpen
  \bibfield  {author} {\bibinfo {author} {\bibfnamefont {K.}~\bibnamefont
  {Gouder}}, \bibinfo {author} {\bibfnamefont {M.}~\bibnamefont {Potter}}, \
  and\ \bibinfo {author} {\bibfnamefont {J.F.}\ \bibnamefont {Morrison}},\
  }\bibfield  {title} {\enquote {\bibinfo {title} {Turbulent friction drag
  reduction using electroactive polymer and electromagnetically driven
  surfaces},}\ }\href@noop {} {\bibfield  {journal} {\bibinfo  {journal} {Exp.
  Fluids}\ }\textbf {\bibinfo {volume} {54}},\ \bibinfo {pages} {1} (\bibinfo
  {year} {2013})}\BibitemShut {NoStop}%
\bibitem [{\citenamefont {Choi}\ \emph {et~al.}(2011)\citenamefont {Choi},
  \citenamefont {Jukes},\ and\ \citenamefont
  {Whalley}}]{choi-jukes-whalley-2011}%
  \BibitemOpen
  \bibfield  {author} {\bibinfo {author} {\bibfnamefont {K.-S.}\ \bibnamefont
  {Choi}}, \bibinfo {author} {\bibfnamefont {T.}~\bibnamefont {Jukes}}, \ and\
  \bibinfo {author} {\bibfnamefont {R.}~\bibnamefont {Whalley}},\ }\bibfield
  {title} {\enquote {\bibinfo {title} {Turbulent boundary-layer control with
  plasma actuators},}\ }\href@noop {} {\bibfield  {journal} {\bibinfo
  {journal} {Phil. Trans. R. Soc.}\ }\textbf {\bibinfo {volume} {369}},\
  \bibinfo {pages} {1443} (\bibinfo {year} {2011})}\BibitemShut {NoStop}%
\bibitem [{\citenamefont {Keefe}(1998)}]{keefe-1998}%
  \BibitemOpen
  \bibfield  {author} {\bibinfo {author} {\bibfnamefont {L.}~\bibnamefont
  {Keefe}},\ }\bibfield  {title} {\enquote {\bibinfo {title} {Method and
  apparatus for reducing the drag of flows over surfaces},}\ }\href@noop {}
  {\bibfield  {journal} {\bibinfo  {journal} {United States Patent}\ }\textbf
  {\bibinfo {volume} {5,803,409}} (\bibinfo {year} {1998})}\BibitemShut
  {NoStop}%
\bibitem [{\citenamefont {Ricco}\ and\ \citenamefont
  {Hahn}(2013)}]{ricco-hahn-2013}%
  \BibitemOpen
  \bibfield  {author} {\bibinfo {author} {\bibfnamefont {P.}~\bibnamefont
  {Ricco}}\ and\ \bibinfo {author} {\bibfnamefont {S.}~\bibnamefont {Hahn}},\
  }\bibfield  {title} {\enquote {\bibinfo {title} {Turbulent drag reduction
  through rotating discs},}\ }\href@noop {} {\bibfield  {journal} {\bibinfo
  {journal} {J. Fluid Mech.}\ }\textbf {\bibinfo {volume} {722}},\ \bibinfo
  {pages} {267} (\bibinfo {year} {2013})}\BibitemShut {NoStop}%
\bibitem [{\citenamefont {Fukagata}\ \emph {et~al.}(2002)\citenamefont
  {Fukagata}, \citenamefont {Iwamoto},\ and\ \citenamefont
  {Kasagi}}]{fukagata-iwamoto-kasagi-2002}%
  \BibitemOpen
  \bibfield  {author} {\bibinfo {author} {\bibfnamefont {K.}~\bibnamefont
  {Fukagata}}, \bibinfo {author} {\bibfnamefont {K.}~\bibnamefont {Iwamoto}}, \
  and\ \bibinfo {author} {\bibfnamefont {N.}~\bibnamefont {Kasagi}},\
  }\bibfield  {title} {\enquote {\bibinfo {title} {Contribution of {R}eynolds
  stress distribution to the skin friction in wall-bounded flows},}\
  }\href@noop {} {\bibfield  {journal} {\bibinfo  {journal} {Phys. Fluids}\
  }\textbf {\bibinfo {volume} {14}},\ \bibinfo {pages} {73} (\bibinfo {year}
  {2002})}\BibitemShut {NoStop}%
\bibitem [{\citenamefont {von K{\'a}rm{\'a}n}(1921)}]{karman-1921}%
  \BibitemOpen
  \bibfield  {author} {\bibinfo {author} {\bibnamefont {von K{\'a}rm{\'a}n}},\
  }\bibfield  {title} {\enquote {\bibinfo {title} {{\"U}ber laminare und
  turbulente reibung},}\ }\href@noop {} {\bibfield  {journal} {\bibinfo
  {journal} {ZAMM-Journal of Applied Mathematics and Mechanics/Zeitschrift
  f{\"u}r Angewandte Mathematik und Mechanik}\ }\textbf {\bibinfo {volume}
  {1}},\ \bibinfo {pages} {233} (\bibinfo {year} {1921})}\BibitemShut {NoStop}%
\bibitem [{\citenamefont {Cochran}(1934)}]{cochran-1934}%
  \BibitemOpen
  \bibfield  {author} {\bibinfo {author} {\bibfnamefont {W.G.}\ \bibnamefont
  {Cochran}},\ }\bibfield  {title} {\enquote {\bibinfo {title} {The flow due to
  a rotating disc},}\ }in\ \href@noop {} {\emph {\bibinfo {booktitle}
  {Mathematical Proceedings of the Cambridge Philosophical Society}}},\
  Vol.~\bibinfo {volume} {30}\ (\bibinfo {organization} {Cambridge Univ
  Press},\ \bibinfo {year} {1934})\ p.\ \bibinfo {pages} {365}\BibitemShut
  {NoStop}%
\bibitem [{\citenamefont {Rogers}\ and\ \citenamefont
  {Lance}(1960)}]{rogers-lance-1960}%
  \BibitemOpen
  \bibfield  {author} {\bibinfo {author} {\bibfnamefont {M.H.}\ \bibnamefont
  {Rogers}}\ and\ \bibinfo {author} {\bibfnamefont {G.N.}\ \bibnamefont
  {Lance}},\ }\bibfield  {title} {\enquote {\bibinfo {title} {The rotationally
  symmetric flow of a viscous fluid in the presence of an infinite rotating
  disk},}\ }\href@noop {} {\bibfield  {journal} {\bibinfo  {journal} {J. Fluid
  Mech.}\ }\textbf {\bibinfo {volume} {7}},\ \bibinfo {pages} {617} (\bibinfo
  {year} {1960})}\BibitemShut {NoStop}%
\bibitem [{\citenamefont {Wang}(1989)}]{wang-1989}%
  \BibitemOpen
  \bibfield  {author} {\bibinfo {author} {\bibfnamefont {C.Y.}\ \bibnamefont
  {Wang}},\ }\bibfield  {title} {\enquote {\bibinfo {title} {Shear flow over a
  rotating plate},}\ }\href@noop {} {\bibfield  {journal} {\bibinfo  {journal}
  {App. Sc. Res.}\ }\textbf {\bibinfo {volume} {46}},\ \bibinfo {pages} {89}
  (\bibinfo {year} {1989})}\BibitemShut {NoStop}%
\bibitem [{\citenamefont {Klewicki}\ and\ \citenamefont
  {Hill}(2003)}]{klewicki-hill-2003}%
  \BibitemOpen
  \bibfield  {author} {\bibinfo {author} {\bibfnamefont {J.C.}\ \bibnamefont
  {Klewicki}}\ and\ \bibinfo {author} {\bibfnamefont {R.B.}\ \bibnamefont
  {Hill}},\ }\bibfield  {title} {\enquote {\bibinfo {title} {Laminar boundary
  layer response to rotation of a finite diameter surface patch},}\ }\href@noop
  {} {\bibfield  {journal} {\bibinfo  {journal} {Phys. Fluids}\ }\textbf
  {\bibinfo {volume} {15}},\ \bibinfo {pages} {101} (\bibinfo {year}
  {2003})}\BibitemShut {NoStop}%
\bibitem [{\citenamefont {Lingwood}(1995)}]{lingwood-1995}%
  \BibitemOpen
  \bibfield  {author} {\bibinfo {author} {\bibfnamefont {R.J.}\ \bibnamefont
  {Lingwood}},\ }\bibfield  {title} {\enquote {\bibinfo {title} {Absolute
  instability of the boundary layer on a rotating disk},}\ }\href@noop {}
  {\bibfield  {journal} {\bibinfo  {journal} {J Fluid Mech.}\ }\textbf
  {\bibinfo {volume} {299}},\ \bibinfo {pages} {17} (\bibinfo {year}
  {1995})}\BibitemShut {NoStop}%
\bibitem [{\citenamefont {Lingwood}(1996)}]{lingwood-1996}%
  \BibitemOpen
  \bibfield  {author} {\bibinfo {author} {\bibfnamefont {RJ}~\bibnamefont
  {Lingwood}},\ }\bibfield  {title} {\enquote {\bibinfo {title} {An
  experimental study of absolute instability of the rotating-disk
  boundary-layer flow},}\ }\href@noop {} {\bibfield  {journal} {\bibinfo
  {journal} {J Fluid Mech.}\ }\textbf {\bibinfo {volume} {314}},\ \bibinfo
  {pages} {373} (\bibinfo {year} {1996})}\BibitemShut {NoStop}%
\bibitem [{\citenamefont {Lingwood}(1997)}]{lingwood-1997b}%
  \BibitemOpen
  \bibfield  {author} {\bibinfo {author} {\bibfnamefont {R.J.}\ \bibnamefont
  {Lingwood}},\ }\bibfield  {title} {\enquote {\bibinfo {title} {Absolute
  instability of the {E}kman layer and related rotating flows},}\ }\href@noop
  {} {\bibfield  {journal} {\bibinfo  {journal} {J Fluid Mech.}\ }\textbf
  {\bibinfo {volume} {331}},\ \bibinfo {pages} {405} (\bibinfo {year}
  {1997})}\BibitemShut {NoStop}%
\bibitem [{\citenamefont {Wise}\ and\ \citenamefont
  {Ricco}(2014)}]{wise-ricco-2014}%
  \BibitemOpen
  \bibfield  {author} {\bibinfo {author} {\bibfnamefont {D.~J.}\ \bibnamefont
  {Wise}}\ and\ \bibinfo {author} {\bibfnamefont {P.}~\bibnamefont {Ricco}},\
  }\bibfield  {title} {\enquote {\bibinfo {title} {Turbulent drag reduction
  through oscillating discs},}\ }\href@noop {} {\bibfield  {journal} {\bibinfo
  {journal} {J. Fluid Mech.}\ }\textbf {\bibinfo {volume} {746}},\ \bibinfo
  {pages} {536} (\bibinfo {year} {2014})}\BibitemShut {NoStop}%
\bibitem [{\citenamefont {Gibson}(2014)}]{channelflow-2006}%
  \BibitemOpen
  \bibfield  {author} {\bibinfo {author} {\bibfnamefont {J.~F.}\ \bibnamefont
  {Gibson}},\ }\href@noop {} {\emph {\bibinfo {title} {{Channelflow}: {A}
  spectral {Navier-Stokes} simulator in {C}++}}},\ \bibinfo {type} {Tech.
  Rep.}\ (\bibinfo  {institution} {U. New Hampshire},\ \bibinfo {year} {2014})\
  \bibinfo {note} {{\tt {Channelflow.org}}}\BibitemShut {NoStop}%
\bibitem [{\citenamefont {Kleiser}\ and\ \citenamefont
  {Schumann}(1980)}]{kleiser-schumann-1980}%
  \BibitemOpen
  \bibfield  {author} {\bibinfo {author} {\bibfnamefont {L.}~\bibnamefont
  {Kleiser}}\ and\ \bibinfo {author} {\bibfnamefont {U.}~\bibnamefont
  {Schumann}},\ }\bibfield  {title} {\enquote {\bibinfo {title} {Treatment of
  incompressibility and boundary conditions in 3-{D} numerical spectral
  simulations of plane channel flows},}\ }in\ \href@noop {} {\emph {\bibinfo
  {booktitle} {Proc. 3rd {GAMM} {C}onf. {N}umerical {M}ethods in {F}luid
  {M}echanics}}},\ \bibinfo {editor} {edited by\ \bibinfo {editor}
  {\bibfnamefont {E.}~\bibnamefont {Hirschel}}},\ \bibinfo {organization}
  {GAMM}\ (\bibinfo  {publisher} {Vieweg},\ \bibinfo {year} {1980})\ p.\
  \bibinfo {pages} {165}\BibitemShut {NoStop}%
\bibitem [{\citenamefont {Canuto}\ \emph {et~al.}(1988)\citenamefont {Canuto},
  \citenamefont {Hussaini}, \citenamefont {Quarteroni},\ and\ \citenamefont
  {Zang}}]{canuto-etal-1988}%
  \BibitemOpen
  \bibfield  {author} {\bibinfo {author} {\bibfnamefont {C.}~\bibnamefont
  {Canuto}}, \bibinfo {author} {\bibfnamefont {M.Y.}\ \bibnamefont {Hussaini}},
  \bibinfo {author} {\bibfnamefont {A.}~\bibnamefont {Quarteroni}}, \ and\
  \bibinfo {author} {\bibfnamefont {T.A.}\ \bibnamefont {Zang}},\ }\href@noop
  {} {\emph {\bibinfo {title} {{S}pectral {M}ethods in {F}luid {D}ynamics}}}\
  (\bibinfo  {publisher} {Springer-Verlag, New York},\ \bibinfo {year}
  {1988})\BibitemShut {NoStop}%
\bibitem [{\citenamefont {Cimarelli}\ \emph {et~al.}(2013)\citenamefont
  {Cimarelli}, \citenamefont {Frohnapfel}, \citenamefont {Hasegawa},
  \citenamefont {De~Angelis},\ and\ \citenamefont
  {Quadrio}}]{cimarelli-etal-2013}%
  \BibitemOpen
  \bibfield  {author} {\bibinfo {author} {\bibfnamefont {A.}~\bibnamefont
  {Cimarelli}}, \bibinfo {author} {\bibfnamefont {B.}~\bibnamefont
  {Frohnapfel}}, \bibinfo {author} {\bibfnamefont {Y.}~\bibnamefont
  {Hasegawa}}, \bibinfo {author} {\bibfnamefont {E.}~\bibnamefont
  {De~Angelis}}, \ and\ \bibinfo {author} {\bibfnamefont {M.}~\bibnamefont
  {Quadrio}},\ }\bibfield  {title} {\enquote {\bibinfo {title} {Prediction of
  turbulence control for arbitrary periodic spanwise wall movement},}\
  }\href@noop {} {\bibfield  {journal} {\bibinfo  {journal} {Phys. Fluids}\
  }\textbf {\bibinfo {volume} {25}} (\bibinfo {year} {2013})}\BibitemShut
  {NoStop}%
\bibitem [{\citenamefont {Zhou}\ and\ \citenamefont
  {Ball}(2008)}]{zhou-ball-2006}%
  \BibitemOpen
  \bibfield  {author} {\bibinfo {author} {\bibfnamefont {D.}~\bibnamefont
  {Zhou}}\ and\ \bibinfo {author} {\bibfnamefont {K.S.}\ \bibnamefont {Ball}},\
  }\bibfield  {title} {\enquote {\bibinfo {title} {Turbulent drag reduction by
  spanwise wall oscillations},}\ }\href@noop {} {\bibfield  {journal} {\bibinfo
   {journal} {Int. J. Eng. Trans. A Basics}\ }\textbf {\bibinfo {volume}
  {21}},\ \bibinfo {pages} {85} (\bibinfo {year} {2008})}\BibitemShut {NoStop}%
\bibitem [{\citenamefont {Batchelor}(1967)}]{batchelor-1967}%
  \BibitemOpen
  \bibfield  {author} {\bibinfo {author} {\bibfnamefont {G.~K.}\ \bibnamefont
  {Batchelor}},\ }\href@noop {} {\emph {\bibinfo {title} {{A}n {I}ntroduction
  to {F}luid {D}ynamics}}}\ (\bibinfo  {publisher} {Cambridge University
  Press},\ \bibinfo {year} {1967})\BibitemShut {NoStop}%
\bibitem [{\citenamefont {Schlichting}(1979)}]{schlichting-1979}%
  \BibitemOpen
  \bibfield  {author} {\bibinfo {author} {\bibfnamefont {H.}~\bibnamefont
  {Schlichting}},\ }\href@noop {} {\emph {\bibinfo {title} {Boundary-layer
  theory}}}\ (\bibinfo  {publisher} {McGraw Hill.Inc.},\ \bibinfo {year}
  {1979})\BibitemShut {NoStop}%
\bibitem [{\citenamefont {Koch}\ and\ \citenamefont
  {Kozulovic}(2013)}]{koch-kozulovic-2013}%
  \BibitemOpen
  \bibfield  {author} {\bibinfo {author} {\bibfnamefont {H.}~\bibnamefont
  {Koch}}\ and\ \bibinfo {author} {\bibfnamefont {D.}~\bibnamefont
  {Kozulovic}},\ }\bibfield  {title} {\enquote {\bibinfo {title} {Drag
  reduction by boundary layer control with passively moving wall},}\ }in\
  \href@noop {} {\emph {\bibinfo {booktitle} {ASME 2013 Fluids Engineering
  Division Summer Meeting}}}\ (\bibinfo {organization} {American Society of
  Mechanical Engineers},\ \bibinfo {year} {2013})\ p.\ \bibinfo {pages}
  {V01BT15A004}\BibitemShut {NoStop}%
\bibitem [{\citenamefont {H{\oe}pffner}\ and\ \citenamefont
  {Fukagata}(2009)}]{hoepffner-fukagata-2009}%
  \BibitemOpen
  \bibfield  {author} {\bibinfo {author} {\bibfnamefont {J.}~\bibnamefont
  {H{\oe}pffner}}\ and\ \bibinfo {author} {\bibfnamefont {K.}~\bibnamefont
  {Fukagata}},\ }\bibfield  {title} {\enquote {\bibinfo {title} {Pumping or
  drag reduction?}}\ }\href@noop {} {\bibfield  {journal} {\bibinfo  {journal}
  {J. Fluid Mech.}\ }\textbf {\bibinfo {volume} {635}},\ \bibinfo {pages} {171}
  (\bibinfo {year} {2009})}\BibitemShut {NoStop}%
\bibitem [{\citenamefont {Sasamori}\ \emph {et~al.}(2014)\citenamefont
  {Sasamori}, \citenamefont {Mamori}, \citenamefont {Iwamoto},\ and\
  \citenamefont {Murata}}]{sasamori-etal-2014}%
  \BibitemOpen
  \bibfield  {author} {\bibinfo {author} {\bibfnamefont {M.}~\bibnamefont
  {Sasamori}}, \bibinfo {author} {\bibfnamefont {H.}~\bibnamefont {Mamori}},
  \bibinfo {author} {\bibfnamefont {K.}~\bibnamefont {Iwamoto}}, \ and\
  \bibinfo {author} {\bibfnamefont {A.}~\bibnamefont {Murata}},\ }\bibfield
  {title} {\enquote {\bibinfo {title} {Experimental study on drag--reduction
  effect due to sinusoidal riblets in turbulent channel flow},}\ }\href@noop {}
  {\bibfield  {journal} {\bibinfo  {journal} {Experiments in Fluids}\ }\textbf
  {\bibinfo {volume} {55}},\ \bibinfo {pages} {1} (\bibinfo {year}
  {2014})}\BibitemShut {NoStop}%
\bibitem [{\citenamefont {Choi}\ \emph {et~al.}(1994)\citenamefont {Choi},
  \citenamefont {Moin},\ and\ \citenamefont {Kim}}]{choi-moin-kim-1994}%
  \BibitemOpen
  \bibfield  {author} {\bibinfo {author} {\bibfnamefont {H.}~\bibnamefont
  {Choi}}, \bibinfo {author} {\bibfnamefont {P.}~\bibnamefont {Moin}}, \ and\
  \bibinfo {author} {\bibfnamefont {J.}~\bibnamefont {Kim}},\ }\bibfield
  {title} {\enquote {\bibinfo {title} {Active turbulence control for drag
  reduction in wall-bounded flows},}\ }\href@noop {} {\bibfield  {journal}
  {\bibinfo  {journal} {J. Fluid Mech.}\ }\textbf {\bibinfo {volume} {262}},\
  \bibinfo {pages} {75} (\bibinfo {year} {1994})}\BibitemShut {NoStop}%
\bibitem [{\citenamefont {Quadrio}\ and\ \citenamefont
  {Ricco}(2004)}]{quadrio-ricco-2004}%
  \BibitemOpen
  \bibfield  {author} {\bibinfo {author} {\bibfnamefont {M.}~\bibnamefont
  {Quadrio}}\ and\ \bibinfo {author} {\bibfnamefont {P.}~\bibnamefont
  {Ricco}},\ }\bibfield  {title} {\enquote {\bibinfo {title} {Critical
  assessment of turbulent drag reduction through spanwise wall oscillations},}\
  }\href@noop {} {\bibfield  {journal} {\bibinfo  {journal} {J. Fluid Mech.}\
  }\textbf {\bibinfo {volume} {521}},\ \bibinfo {pages} {251} (\bibinfo {year}
  {2004})}\BibitemShut {NoStop}%
\bibitem [{\citenamefont {Hinze}(1975)}]{hinze-1975}%
  \BibitemOpen
  \bibfield  {author} {\bibinfo {author} {\bibfnamefont {J.O.}\ \bibnamefont
  {Hinze}},\ }\href@noop {} {\emph {\bibinfo {title} {Turbulence}}}\ (\bibinfo
  {publisher} {McGraw Hill, Inc. -- Second Edition},\ \bibinfo {year}
  {1975})\BibitemShut {NoStop}%
\bibitem [{\citenamefont {Rosenblat}(1959)}]{rosenblat-1959}%
  \BibitemOpen
  \bibfield  {author} {\bibinfo {author} {\bibfnamefont {S.}~\bibnamefont
  {Rosenblat}},\ }\bibfield  {title} {\enquote {\bibinfo {title} {Torsional
  oscillations of a plane in a viscous fluid},}\ }\href@noop {} {\bibfield
  {journal} {\bibinfo  {journal} {J. Fluid Mech.}\ }\textbf {\bibinfo {volume}
  {6}},\ \bibinfo {pages} {206} (\bibinfo {year} {1959})}\BibitemShut {NoStop}%
\end{thebibliography}%

\begin{figure}
 \centering 
 \includegraphics[width=0.99\textwidth]{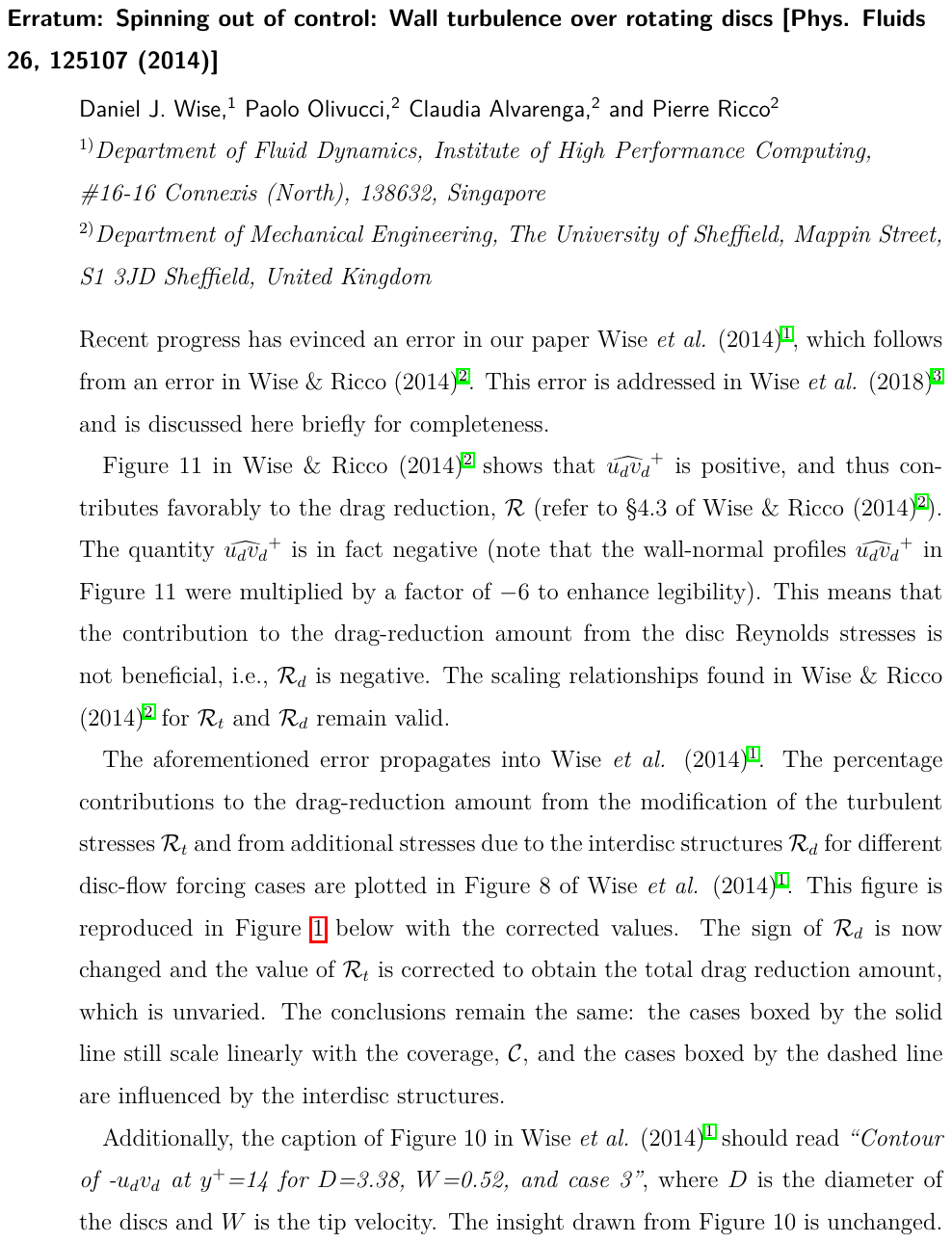}
\end{figure}

\begin{figure}
 \centering 
 \includegraphics[width=0.99\textwidth]{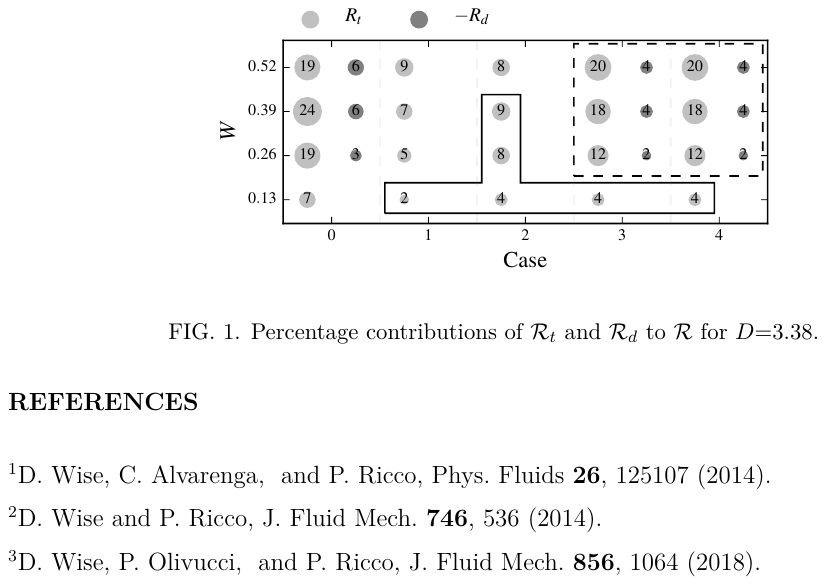}
\end{figure}

\end{document}